\documentclass[aps,prd,twocolumn,showpacs,superscriptaddress,groupedaddress,floatfix]{revtex4}  

  \usepackage{graphicx}  \usepackage{dcolumn}   \usepackage{bm}        \usepackage{amssymb}   \usepackage{amsmath}
\usepackage{footnote}
\usepackage{setspace}
\usepackage{enumerate}
\usepackage{color}

\usepackage{amssymb,amsmath,mathrsfs}

\usepackage{dcolumn}\newcolumntype{.}{D{.}{.}{-1}}
\newcolumntype{,}{D{,}{\relax}{-1}}
\newcolumntype{+}{D{+}{\,\pm\,}{-1}}
\newcolumntype{^}{D{^}{^}{-1}}

\begin{document}

\hspace{5.2in} \mbox{FERMILAB-PUB-15-312-E}

\makeatletter{}
\def\sqrts{\ensuremath{\sqrt s}}
\newcommand{\invisible}[1]{}
\newcommand{\pt} 	{\mbox{$p_T$}}
\newcommand{\dzero}     {D0} \newcommand{\ttbar}     {\ensuremath{t\bar{t}}}
\newcommand{\sigmattbar} {\ensuremath{\sigma_{t\bar{t}}}}
\newcommand{\ppbar}     {\ensuremath{p\bar{p}}}
\newcommand{\qqbar}     {\ensuremath{q\bar{q}}}
\newcommand{\herwig}    {\textsc{herwig}}
\newcommand{\pythia}    {\textsc{pythia}}
\newcommand{\alpgen}    {\textsc{alpgen}}
\newcommand{\evtgen}    {\textsc{evtgen}}
\newcommand{\tauola}    {\textsc{tauola}}
\newcommand{\geant}     {\textsc{geant}}
\newcommand{\mcatnlo}   {\textsc{mc@nlo}}
\newcommand{\madgraph}   {\textsc{madgraph}} \hyphenation{madgraph}
\newcommand{\zgrad}     {\textsc{zgrad}}
\newcommand{\met}       {\mbox{$\not\!\!E_T$}}
\newcommand{\vegas}{{\sc vegas}}

\newcommand{\metsig}       {\mbox{$\not\!\!E_T^{\text{sig}}$}}
\newcommand{\metcal}    {\mbox{$\not\!\!E_{Tcal}$}}
\newcommand{\rar}       {\rightarrow}
\newcommand{\eps}       {\epsilon}
\newcommand{\ztau}      {$Z\to\tau\tau$}
\newcommand{\zmu}       {$Z\to\mu\mu$}
\newcommand{\ze}        {$Z\to ee$}
\newcommand{\zl} 	{$Z \to \ell \ell$}

\newcommand{\defAll}    {\ensuremath{\All = \frac{N(\Delta\eta > 0) - N(\Delta\eta < 0)}{N(\Delta\eta > 0) + N(\Delta\eta < 0)}}}
\newcommand{\Alfb}      {\ensuremath{A^{\ell}_{\rm FB}}}
\newcommand{\defAlfb}   {\ensuremath{\Alfb = \frac{N_\ell(q\times \eta>0) - N_\ell(q\times \eta<0)}{N_\ell(q\times \eta>0) + N_\ell(q\times \eta<0)}}}
\newcommand{\Alfbp}			{\ensuremath{A^{\ell^+}_{\rm FB}}} 
\newcommand{\defAlfbp}   {\ensuremath{\Alfbp = \frac{N_{\ell^+}(\eta>0) - N_{\ell^+}(\eta<0)}{N_{\ell^+}(\eta>0) + N_{\ell^+}(\eta<0)}}}
\newcommand{\Alfbn}			{\ensuremath{A^{\ell^-}_{\rm FB}}}
\newcommand{\defAlfbn}   {\ensuremath{\Alfbn = \frac{N_{\ell^-}(\eta>0) - N_{\ell^-}(\eta<0)}{N_{\ell^-}(\eta>0) + N_{\ell^-}(\eta<0)}}}
\newcommand{\All}       {\ensuremath{A^{\ell\ell}_{\rm FB}}}

\newcommand{\Attfb}       {\ensuremath{A^{t\bar t}}}
\newcommand{\Att}       {\Attfb}
\newcommand{\Attraw}          {\ensuremath{A^{t\bar t}_{\rm raw}}}
\newcommand{\Adataraw}          {\ensuremath{A^{\rm data}_{\rm raw}}}
\newcommand{\Abkgraw}          {\ensuremath{A^{\rm bkg}_{\rm raw}}}

\newcommand{\Adataminusbkgraw}          {\ensuremath{A^{{\rm data}-{\rm bkg}}_{\rm raw}}}

\def\ddif{ {d}}
\newcommand{\AP}[2]     {\ensuremath{A^{\ell ^{#1}}_{#2}}}
\newcommand{\AAP}[3]     {\ensuremath{A^{\ell ^{#1}{#2} }_{#3}}}
\newcommand{\Polar}[2]     {\ensuremath{P^{#1}_{#2}}}

\newcommand{\defAP}[2]       {\AP{#1}{#2}=\frac{N(\cos\theta^{#1}>0) - N(\cos\theta^{#1}<0)}{N(\cos\theta^{#1}>0) + N(\cos\theta^{#1}<0)}}
\newcommand{\APraw}{\ensuremath{\frac 1 2 \left(\AP{+}{\rm beam, raw} -\AP{-}{\invisible{\rm beam,} raw}\right)}}

\newcommand{\Ptt}{\ensuremath{\kappa\Polar{}{\invisible{\rm beam}}}}
\newcommand{\Praw}{\ensuremath{\kappa\Polar{}{\invisible{\rm beam,} \rm raw}}}
\newcommand{\Pttraw}{\ensuremath{\kappa\Polar{}{\invisible{\rm beam,} \rm raw}}}

\newcommand{\ie}{{\it i.e.}}
\def\eg{{\it e.g.}}

\def\yt {\ensuremath{ y_t}}
\def\ytbar {\ensuremath{ y_{\bar t}}}
\def\dyttbar{\ensuremath{\Delta y_{\ttbar}}}
\def\gev {\ensuremath{\mathrm{GeV}}}
\def\tev {\ensuremath{\mathrm{TeV}}}

\newcommand{\defAtt}{\ensuremath{\Att=   \frac {  N ( \dyttbar>0) -  N ( \dyttbar <0) }{  N ( \dyttbar>0) + N ( \dyttbar <0) }}}

\def \ee{\ensuremath{ee}}
\def \emu{\ensuremath{e\mu}}
\def \mumu{\ensuremath{\mu\mu}}

\newcommand{\lumi}     {$\mathrm{9.7~fb^{-1}}$}
\newcommand{\lumiprev}    {$\mathrm{5.4~fb^{-1}}$}

\newcommand{\doublet}[2]{{\left(\begin{array}{c} #1 \\ {#2} \end{array}\right)}}
\newcommand{\matrixfour}[4]{{\left(\begin{array}{cc} {#1} & {#2}\\ {#3} &{#4} \end{array}\right)}}

\title{
Simultaneous measurement of
forward-backward asymmetry and top polarization
in dilepton final states
from \ttbar\ production at the Tevatron
}

\makeatletter{}\affiliation{LAFEX, Centro Brasileiro de Pesquisas F\'{i}sicas, Rio de Janeiro, Brazil}
\affiliation{Universidade do Estado do Rio de Janeiro, Rio de Janeiro, Brazil}
\affiliation{Universidade Federal do ABC, Santo Andr\'e, Brazil}
\affiliation{University of Science and Technology of China, Hefei, People's Republic of China}
\affiliation{Universidad de los Andes, Bogot\'a, Colombia}
\affiliation{Charles University, Faculty of Mathematics and Physics, Center for Particle Physics, Prague, Czech Republic}
\affiliation{Czech Technical University in Prague, Prague, Czech Republic}
\affiliation{Institute of Physics, Academy of Sciences of the Czech Republic, Prague, Czech Republic}
\affiliation{Universidad San Francisco de Quito, Quito, Ecuador}
\affiliation{LPC, Universit\'e Blaise Pascal, CNRS/IN2P3, Clermont, France}
\affiliation{LPSC, Universit\'e Joseph Fourier Grenoble 1, CNRS/IN2P3, Institut National Polytechnique de Grenoble, Grenoble, France}
\affiliation{CPPM, Aix-Marseille Universit\'e, CNRS/IN2P3, Marseille, France}
\affiliation{LAL, Universit\'e Paris-Sud, CNRS/IN2P3, Orsay, France}
\affiliation{LPNHE, Universit\'es Paris VI and VII, CNRS/IN2P3, Paris, France}
\affiliation{CEA, Irfu, SPP, Saclay, France}
\affiliation{IPHC, Universit\'e de Strasbourg, CNRS/IN2P3, Strasbourg, France}
\affiliation{IPNL, Universit\'e Lyon 1, CNRS/IN2P3, Villeurbanne, France and Universit\'e de Lyon, Lyon, France}
\affiliation{III. Physikalisches Institut A, RWTH Aachen University, Aachen, Germany}
\affiliation{Physikalisches Institut, Universit\"at Freiburg, Freiburg, Germany}
\affiliation{II. Physikalisches Institut, Georg-August-Universit\"at G\"ottingen, G\"ottingen, Germany}
\affiliation{Institut f\"ur Physik, Universit\"at Mainz, Mainz, Germany}
\affiliation{Ludwig-Maximilians-Universit\"at M\"unchen, M\"unchen, Germany}
\affiliation{Panjab University, Chandigarh, India}
\affiliation{Delhi University, Delhi, India}
\affiliation{Tata Institute of Fundamental Research, Mumbai, India}
\affiliation{University College Dublin, Dublin, Ireland}
\affiliation{Korea Detector Laboratory, Korea University, Seoul, Korea}
\affiliation{CINVESTAV, Mexico City, Mexico}
\affiliation{Nikhef, Science Park, Amsterdam, the Netherlands}
\affiliation{Radboud University Nijmegen, Nijmegen, the Netherlands}
\affiliation{Joint Institute for Nuclear Research, Dubna, Russia}
\affiliation{Institute for Theoretical and Experimental Physics, Moscow, Russia}
\affiliation{Moscow State University, Moscow, Russia}
\affiliation{Institute for High Energy Physics, Protvino, Russia}
\affiliation{Petersburg Nuclear Physics Institute, St. Petersburg, Russia}
\affiliation{Instituci\'{o} Catalana de Recerca i Estudis Avan\c{c}ats (ICREA) and Institut de F\'{i}sica d'Altes Energies (IFAE), Barcelona, Spain}
\affiliation{Uppsala University, Uppsala, Sweden}
\affiliation{Taras Shevchenko National University of Kyiv, Kiev, Ukraine}
\affiliation{Lancaster University, Lancaster LA1 4YB, United Kingdom}
\affiliation{Imperial College London, London SW7 2AZ, United Kingdom}
\affiliation{The University of Manchester, Manchester M13 9PL, United Kingdom}
\affiliation{University of Arizona, Tucson, Arizona 85721, USA}
\affiliation{University of California Riverside, Riverside, California 92521, USA}
\affiliation{Florida State University, Tallahassee, Florida 32306, USA}
\affiliation{Fermi National Accelerator Laboratory, Batavia, Illinois 60510, USA}
\affiliation{University of Illinois at Chicago, Chicago, Illinois 60607, USA}
\affiliation{Northern Illinois University, DeKalb, Illinois 60115, USA}
\affiliation{Northwestern University, Evanston, Illinois 60208, USA}
\affiliation{Indiana University, Bloomington, Indiana 47405, USA}
\affiliation{Purdue University Calumet, Hammond, Indiana 46323, USA}
\affiliation{University of Notre Dame, Notre Dame, Indiana 46556, USA}
\affiliation{Iowa State University, Ames, Iowa 50011, USA}
\affiliation{University of Kansas, Lawrence, Kansas 66045, USA}
\affiliation{Louisiana Tech University, Ruston, Louisiana 71272, USA}
\affiliation{Northeastern University, Boston, Massachusetts 02115, USA}
\affiliation{University of Michigan, Ann Arbor, Michigan 48109, USA}
\affiliation{Michigan State University, East Lansing, Michigan 48824, USA}
\affiliation{University of Mississippi, University, Mississippi 38677, USA}
\affiliation{University of Nebraska, Lincoln, Nebraska 68588, USA}
\affiliation{Rutgers University, Piscataway, New Jersey 08855, USA}
\affiliation{Princeton University, Princeton, New Jersey 08544, USA}
\affiliation{State University of New York, Buffalo, New York 14260, USA}
\affiliation{University of Rochester, Rochester, New York 14627, USA}
\affiliation{State University of New York, Stony Brook, New York 11794, USA}
\affiliation{Brookhaven National Laboratory, Upton, New York 11973, USA}
\affiliation{Langston University, Langston, Oklahoma 73050, USA}
\affiliation{University of Oklahoma, Norman, Oklahoma 73019, USA}
\affiliation{Oklahoma State University, Stillwater, Oklahoma 74078, USA}
\affiliation{Brown University, Providence, Rhode Island 02912, USA}
\affiliation{University of Texas, Arlington, Texas 76019, USA}
\affiliation{Southern Methodist University, Dallas, Texas 75275, USA}
\affiliation{Rice University, Houston, Texas 77005, USA}
\affiliation{University of Virginia, Charlottesville, Virginia 22904, USA}
\affiliation{University of Washington, Seattle, Washington 98195, USA}
\author{V.M.~Abazov} \affiliation{Joint Institute for Nuclear Research, Dubna, Russia}
\author{B.~Abbott} \affiliation{University of Oklahoma, Norman, Oklahoma 73019, USA}
\author{B.S.~Acharya} \affiliation{Tata Institute of Fundamental Research, Mumbai, India}
\author{M.~Adams} \affiliation{University of Illinois at Chicago, Chicago, Illinois 60607, USA}
\author{T.~Adams} \affiliation{Florida State University, Tallahassee, Florida 32306, USA}
\author{J.P.~Agnew} \affiliation{The University of Manchester, Manchester M13 9PL, United Kingdom}
\author{G.D.~Alexeev} \affiliation{Joint Institute for Nuclear Research, Dubna, Russia}
\author{G.~Alkhazov} \affiliation{Petersburg Nuclear Physics Institute, St. Petersburg, Russia}
\author{A.~Alton$^{a}$} \affiliation{University of Michigan, Ann Arbor, Michigan 48109, USA}
\author{A.~Askew} \affiliation{Florida State University, Tallahassee, Florida 32306, USA}
% ...
\collaboration{The D0 Collaboration\footnote{with visitors from
$^{a}$Augustana College, Sioux Falls, SD, USA,
$^{b}$The University of Liverpool, Liverpool, UK,
$^{c}$DESY, Hamburg, Germany,
$^{d}$CONACyT, Mexico City, Mexico,
$^{e}$SLAC, Menlo Park, CA, USA,
$^{f}$University College London, London, UK,
$^{g}$Centro de Investigacion en Computacion - IPN, Mexico City, Mexico,
$^{h}$Universidade Estadual Paulista, S\~ao Paulo, Brazil,
$^{i}$Karlsruher Institut f\"ur Technologie (KIT) - Steinbuch Centre for Computing (SCC),
D-76128 Karlsruhe, Germany,
$^{j}$Office of Science, U.S. Department of Energy, Washington, D.C. 20585, USA,
$^{k}$American Association for the Advancement of Science, Washington, D.C. 20005, USA,
$^{l}$Kiev Institute for Nuclear Research, Kiev, Ukraine,
$^{m}$University of Maryland, College Park, Maryland 20742, USA
and
$^{n}$European Orgnaization for Nuclear Research (CERN), Geneva, Switzerland
}} \noaffiliation
\vskip 0.25cm
      
                                                                                       \date{July 21, 2015}

\begin{abstract}
 \makeatletter{}
We present a simultaneous measurement of  the forward-backward asymmetry and  the top-quark  polarization  in  \ttbar\ production  in dilepton final states using \lumi\ of
proton-antiproton collisions at $\sqrt{s}=1.96$~TeV with the D0 detector.
To reconstruct the distributions of kinematic observables we employ a matrix element technique that calculates 
the likelihood of the possible $\ttbar$ kinematic configurations.
After accounting for the presence of background events and for calibration effects, we obtain a forward-backward asymmetry of
$\Att = (15.0 \pm  6.4 \text{ (stat)} \pm 4.9 \text{ (syst)})\%$ and
a top-quark polarization times spin analyzing power in the beam basis of
$\Ptt = (7.2  \pm 10.5 \text{ (stat)} \pm 4.2 \text{ (syst)})\%$,
with a correlation of $-56\%$ between the  measurements. If we constrain the  forward-backward asymmetry to its expected standard model value,
we obtain a measurement of the top polarization of
$\Ptt = (11.3 \pm 9.1 \text{ (stat)}  \pm 1.9 \text{ (syst)})\%.$
If we constrain the top polarization to its expected standard model value,
we measure a forward-backward asymmetry of
$\Att = (17.5 \pm 5.6 \text{ (stat)}  \pm 3.1 \text{ (syst)})\%.$
A combination with the \dzero\ $\Att$ measurement in the lepton+jets final state yields an asymmetry of
$
\Att = (11.8 \pm 2.5 \text{ (stat)}  \pm 1.3 \text{ (syst)})\% .
$
Within their respective  uncertainties, all  these results 
are consistent with the standard model expectations.

\end{abstract}


\maketitle

\makeatletter{}
\section{Introduction}
\label{sec:introduction}

In proton-antiproton collisions at \sqrts=1.96~TeV,
top quark pairs are predominantly produced in valence quark-antiquark annihilations.
The standard model (SM) predicts this
process to be slightly forward-backward asymmetric: the
top quark (antiquark) tends to be emitted in the same direction as the incoming quark (antiquark), and thus,
in the same direction as the incoming proton (antiproton).
The forward-backward asymmetry in the production is mainly due to  positive contributions from the interference between tree-level  and next-to-leading-order (NLO) box diagrams. It receives smaller negative
contributions from the interference between initial and final state radiation. The interferences with electroweak processes increase the asymmetry. 
In the SM, the asymmetry is predicted to be $\approx 10\%$~\cite{Bernreuther:2012ny,Czakon:2014xsa,Kidonakis:2015ona}.
Within the SM, the longitudinal polarizations of the top quark and antiquark
are due to parity violating electroweak contributions to the production process. The polarization is expected to be
$<0.5\%$ for all choices of the spin quantization axis~\cite{Bernreuther:2006vg,Bernreuther:2008ju}.

Physics beyond the SM could affect the \ttbar\ production mechanism and thus both the forward-backward asymmetry and the top quark and antiquark polarizations.
In particular, models with a new  parity violating interaction such as models with axigluons~\cite{Frampton1987157,axi2,axi3,Frampton:2009rk}, can induce a large positive or negative asymmetry together with a sizable polarization.

The \ttbar\ production asymmetry, \Att, is defined in terms of the difference between the rapidities of the top and antitop quarks, $\dyttbar=\yt-\ytbar$:
\begin{equation}\defAtt,\end{equation} where 
$N(X)$ is the number of events in configuration $X$.
By definition, $\Att$ is independent of effects from the top quark decay such as 
top quark polarization. However, it  requires  the reconstruction of the $\ttbar$ initial state from the decay products, which is challenging especially in dilepton channels.

Measurements of $\Att$ have been performed in the lepton+jets channels  by the CDF~\cite{Aaltonen:2012it} and \dzero~\cite{Abazov:2014cca} Collaborations.
Other 
asymmetry measurements have been performed using observables
based on the  pseudo-rapidity of the leptons from $t\to\ W b\to  \ell \nu b$ decays~\cite{Aaltonen:2013vaf,Aaltonen:2014eva,Abazov:2014oea,Abazov:2013wxa}.
All these measurements agree with the SM predictions.
A comprehensive review of asymmetry measurements performed at the Tevatron can be found in Ref.~\cite{Aguilar-Saavedra:2014kpa}.

As top quarks decay before they hadronize, their spin properties are transferred to the decay products.
The top (antitop) polarization
$\Polar{+}{\hat{n}}$
($\Polar{-}{\hat{n}}$) along a given quantization axis $\hat{n}$
impacts the angular distribution of the positively (negatively) charged lepton~\cite{Bernreuther:2008ju}
\begin{equation} \label{eq:single_lepton_polar}
\frac{\ddif \sigma}
{\ddif \cos\theta^{\pm}}=
\frac 1 2 \left( 
1+ \kappa^{\pm} \Polar{\pm}{\hat{n}}
\cos\theta^{\pm}
\right),
\end{equation}
where  $\theta^{+}$  ($\theta^{-}$) is  the angle between  the positively (negatively) charged lepton  in the top (antitop) rest frame and the quantization axis $\hat{n}$,
and $\kappa^{+}$ ($\kappa^{-}$) is the spin analyzing power of the positively (negatively) charged lepton, which is close to 1  ($-1$) at the 0.1\% level within the SM~\cite{Bernreuther:2008ju}. 
The polarization terms
$\kappa^{+}\Polar{+}{\hat{n}}$  ($\kappa^{-}\Polar{-}{\hat{n}}$) can be obtained as two times the asymmetry of the $\cos\theta^{+}$ ($\cos\theta^{-}$) distribution
\begin{equation}\defAP{\pm}{\hat{n}}.\end{equation}
In the following we use the  beam basis, where $\hat{n}$ is the direction of the proton beam in the \ttbar\ zero momentum frame.
Since we only use the beam basis,
we omit the subscript $\hat{n}$ in the following and
define the  polarization observable as:
\begin{equation}
\Ptt=\frac 1 2 \left( \kappa^{+}\Polar{+}{\invisible{\rm beam}} - \kappa^{-}\Polar{-}{\invisible{\rm beam}}\right)
=\AP{+}{\invisible{\rm beam}}-\AP{-}{\invisible{\rm beam}}.
\end{equation}

Polarization effects have been studied  at the Tevatron in the context of the measurements of the leptonic asymmetries in Ref.~\cite{Abazov:2012oxa},
but no actual measurement of the polarization has been performed.
Measurements of the polarization have been conducted for  top pair production in $pp$ collisions at the Large Hadron Collider
at $\sqrts=7$~TeV. These measurements, performed in different basis choices, are all consistent with the SM expectations~\cite{Chatrchyan:2013wua,Aad:2013ksa}.

This article presents a simultaneous measurement of \Att\ and $\Ptt$ with the \dzero\ detector  in the dilepton decay channel.
It is based on the  full Tevatron integrated luminosity of \lumi\ using  \ttbar\ final states with two leptons, \ee, \emu, or \mumu.
We first reconstruct the \dyttbar\ and  $\cos\theta^{\pm}$ distributions employing
a matrix element integration technique similar to that used for the top-quark mass measurement in the dilepton channel~\cite{Abazov:2011fc}.
These distributions are used to extract  raw measurements of asymmetry and polarization, \Attraw\ and \Praw, in  data.
The experimental observables \Attraw\ and \Praw\  are  correlated
because of acceptance and resolution effects.
Using a \mcatnlo~\cite{Frixione:2002ik, Frixione:2008ym} simulation, we compute the relation between the raw measurements \Attraw\ and \Praw,  and the true parton-level asymmetry and polarization to determine calibration corrections.
We then  extract the final measured values of \Att\ and \Ptt.
This is the first measurement  of the \ttbar\ forward-backward asymmetry
obtained from the reconstructed \dyttbar\ distribution
in the dilepton channel
and the first measurement of the top quark polarization at the Fermilab Tevatron collider.

\makeatletter{}
\section{Detector and object reconstruction}
\label{sec:detector}

The \dzero\ detector used for the Run~II of the Fermilab Tevatron collider is described in detail in Refs.~\cite{run1det,run2det,Abolins2008,Angstadt2010}.
The innermost part of the detector is composed of
a central tracking system with a silicon microstrip tracker (SMT) and
a central fiber tracker embedded within a 2~T solenoidal
magnet.  The tracking system is surrounded by a central preshower
detector and a liquid-argon/uranium calorimeter with
electromagnetic, fine hadronic, and coarse hadronic sections. The 
central calorimeter (CC) covers pseudorapidities~\cite{Note1} of
$|\eta|\lesssim 1.1$.
Two end calorimeters (EC) extend the coverage to $ |\eta| \lesssim 4.2$, while
the coverage of the  pseudorapidity region $1.1\leq|\eta|\leq 1.5$, where the EC and CC overlap, is augmented with  scintillating tiles.
A muon
spectrometer, with pseudorapidity coverage of $|\eta|\lesssim 2$,
is located outside the calorimetry and  comprises  drift
tubes and scintillation counters, before and after iron toroidal magnets.
Trigger decisions are based
on information from the tracking detectors,
calorimeters, and muon spectrometer.

Electrons are reconstructed as isolated clusters in the electromagnetic calorimeter
and required to spatially match a track
in the central tracking system.  
They have to pass a boosted decision tree~\cite{NIM_emid}
criterion based on calorimeter shower shape observables, calorimeter isolation,
a spatial track match probability estimate, and the ratio of the electron cluster energy to track momentum ($E/p$).
Electrons are required to be in the acceptance
of the electromagnetic calorimeter ($|\eta|<1.1$ or $1.5<|\eta|<2.5$).

Muons are identified by the presence of at least one track segment
reconstructed in the acceptance ($|\eta| < 2.0$) of the muon spectrometer that is spatially consistent
with a track in the central tracking detector~\cite{NIM_muonid}. The transverse momentum and charge are
measured by the curvature in the central tracking system.
The angular distance to the nearest jet, 
the momenta of charged particles in a cone around the muon track,
and the energy deposited around the muon trajectory in the calorimeter, are
used to select isolated muons.

Jets are reconstructed from energy deposits in the calorimeter using
an iterative midpoint cone algorithm~\cite{Blazey:2000qt} with a cone radius of ${\cal R}= 0.5$~\cite{Note2}. The jet energies are calibrated using transverse momentum balance in $\gamma+$jet
events~\cite{Abazov:2013hda}.

\section{Dataset and Event Selection}
\label{sec:selection}

The signature of \ttbar\ production in dilepton final states
consists of two high-\pt\ leptons  (electrons or muons), two high-\pt\ jets arising from the showering of two  $b$ quarks,
and  missing transverse energy (\met) due to the undetected neutrinos.
The main backgrounds in this final state arise from 
\zl, with $\ell=e$, $\mu$, or $\tau$,  and diboson production ($WW$, $WZ$, $ZZ$).
These backgrounds are evaluated from Monte Carlo (MC) simulated samples as described in section~\ref{sec:bkgmc}.
Another source of background comes from  $W$+jets and multijet events,
if one or two jets 
are misreconstructed as  electrons or if a muon
from a jet passes the isolation criteria.
The contribution from these backgrounds, denoted as ``instrumental background events'', are estimated directly from data as described in section~\ref{sec:bkgdata}.
Each of the dilepton channels is subject to a different mixture and level of background contamination,
in particular for the background arising from the \zl\ process. 
We therefore apply  slightly different selection requirements.
The  main selection criteria to obtain the final samples of $\ttbar$ candidate events are: \begin{enumerate}

\item We  select two high \pt\ ($\pt >15$~\gev) isolated leptons of opposite charge. 

\item We require that at least one electron passes a single electron trigger condition in the \ee\ channel ($\approx 100\%$ efficient),
and that at least one muon passes a  single muon trigger condition in the \mumu\ channel ($\approx 85\%$ efficient).
In the  $\emu$ channel,
we do not require any specific trigger condition, \ie, we use all \dzero\ trigger terms ($\approx 100\%$ efficient).

\item We require  two or more jets of $\pt>20\ \gev$ and $|\eta|<2.5$.

\item We further improve the purity of the selection by exploiting the significant imbalance of transverse energy due to undetected neutrinos and
by exploiting 
several topological variables:
\begin{enumerate}[(i)]
\item The missing transverse energy \met\
is the magnitude of the missing transverse momentum, 
obtained from the vector sum of the transverse
components of energy deposits in the calorimeter, corrected for
the differences in detector response of the reconstructed muons, electrons, and jets. 
\item
The missing transverse energy significance, \metsig,
is  the logarithm of the probability to measure \met\ under the hypothesis that the true missing transverse momentum is zero,
accounting for the energy resolution of individual reconstructed objects and underlying event~\cite{metsig_thesis}.
\item
$H_T$ is the scalar sum of transverse momenta of the leading lepton and the two leading jets.
\end{enumerate}
In the \ee\ channel
we require 
$\metsig\ge 5$, in the \emu\ channel
$H_T>110$~\gev, and
in the \mumu\ channel
$\metsig\ge 5$ and $\met>40~\gev$.

\item 
We require that at least one of the two leading jets  be $b$-tagged,
using a cut on the
multivariate discriminant described in Ref.~\cite{NIM_btaging}.
The requirement is optimized separately for each channel.
The \ttbar\ selection  efficiencies for these requirements are $\approx 82\%$,
$ \approx 83\%$, and 
$ \approx 75\%$
for  the \ee, \emu, and \mumu\ channels, respectively.

\item 
\label{sec:vegas_cut}
The  integration of the matrix elements by \vegas, described in section~\ref{sec:vegas_implementation}, may return a tiny probability if the event is not consistent with the \ttbar\ event hypothesis 
due to
numerical instabilities in the integration process.
After removing low probability events, we retain signal events in the MC simulation with an efficiency of 99.97\%.  For background MC, the efficiency is $>99.3\%$. 
We remove no data events with this requirement.

\end{enumerate}

\section{Signal and background samples}
\label{sec:signall_bsm_background}

\subsection{Signal}
\label{sec:signal_generation}
To simulate the $\ttbar$ signal, we employ MC events generated with 
 the {\textsc{CTEQ6M1}}  parton distribution functions~(PDFs)~\cite{Nadolsky:2008zw} and 
\mcatnlo~3.4~\cite{Frixione:2002ik, Frixione:2008ym} interfaced to \herwig~6.510~\cite{Corcella:2000bw} for showering and hadronization. 
Alternate signal MC samples are generated to study systematic uncertainties and the shape of the \dyttbar\ distribution. We use a sample generated with 
\alpgen~\cite{alpgen} interfaced to \pythia~6.4~\cite{pythia} for showering and hadronization and  a sample generated with \alpgen\ interfaced to  \herwig~6.510. For both samples we use  \textsc{cteq6l1} PDFs~\cite{Nadolsky:2008zw}.

The \mcatnlo\ generator  is used for the nominal signal sample as it
  simulates NLO effects yielding non-zero \Att.
The value of \Att\ at parton level  without applying any selection requirement is $\Att=(5.23\pm 0.07\text{ (stat)})\%$, which is smaller than a SM prediction~\cite{Czakon:2014xsa} 
that includes higher order effects.
 
The MC events are generated 
with a top-quark mass of $m_t= 172.5~\gev$.
They are normalized to a \ttbar\ production 
cross section of 7.45~pb, which corresponds to the calculation of Ref.~\cite{Moch:2008qy} for $m_t= 172.5~\gev$. 
The generated top mass of 172.5~\gev\ differs from the Tevatron
average mass of $173.18\pm0.94~\gev$~\cite{Aaltonen:2012ra}.
We correct  for this small difference in section~\ref{sec:top_mass_correction}.

\subsection{Beyond standard model benchmarks}
\label{sec:BSM_description}

We also study the five benchmark axigluons models proposed in Ref.~\cite{Carmona:2014gra} that modify  $\ttbar$ production.
For each of the proposed beyond standard model (BSM) benchmarks,
we  produce a \ttbar\ MC sample 
using the \madgraph~\cite{Alwall:2014hca} generator interfaced to \pythia~6.4 for showering and hadronization, and the \textsc{cteq6l1} PDFs.
The $Z'$ boson model proposed in  Ref.~\cite{Carmona:2014gra} is not
considered here since it is  excluded by our \ttbar\ differential cross-section measurement~\cite{Abazov:2014vga}.

\subsection{Background estimated with simulated events}
\label{sec:bkgmc}
The background samples are generated using the {\textsc{CTEQ6L1}}  PDFs.
The \zl\ events are generated using   \alpgen\ interfaced to \pythia~6.4. We normalize the $Z \rightarrow \ell\ell$ sample to the NNLO cross section~\cite{Gavin:2010az}.
The $p_{T}$ distribution of $Z$ bosons is weighted to match the distribution observed in
data~\cite{Abazov:2007ac}, taking into account its dependence on the number of reconstructed jets. 
The diboson backgrounds are simulated using {\pythia} and are  normalized to the NLO cross section calculation performed by {\textsc{MCFM}}~\cite{Campbell:1999ah,mcfm}.

\subsection{\dzero\ simulation}
\label{sec:simulation}
The signal and background processes except instrumental background are simulated with
a detailed {\sc geant3}-based~\cite{bib:geant} MC  simulation of the \dzero\
detector. They are processed
with the same reconstruction software as used for  data.
In order to model the effects of multiple $p\bar{p}$
interactions, the MC events are overlaid with events from random
$p\bar{p}$ collisions with the same luminosity distribution as data. 
The jet energy calibration is adjusted in simulated events to match the one measured in data.
Corrections for residual differences between
data and simulation are applied to electrons, muons, and
jets for both  identification efficiencies and energy resolutions.

\subsection{Instrumental background estimated with data}
\label{sec:bkgdata}

The normalization of events
with jets misidentified as electrons is estimated using the ``matrix method"~\cite{Abazov:2007kg} separately for the \ee\ and \emu\ channels.
The contribution from jets producing  identified muons in the \mumu\ channel is obtained using the same selection criteria as for the sample of $\ttbar$ candidate events, but demanding that the leptons have the same charge. In the $\emu$ channel, it is obtained in the same way but after subtracting  the contribution from  events with jets misidentified as electrons.

Once the absolute contribution of instrumental background events has been determined,
we also need ``template samples'' that model their kinematic properties.
In the \emu\ channel, the template for instrumental background events is obtained 
with 
the same selection criteria as for the samples of $\ttbar$ candidate events,
but without applying the complete set of electron selection criteria.
For the \mumu\ and \ee\ channels, the  contributions from instrumental background events is negligible and the result is not sensitive to the choice of template.
For simplicity, we re-employ the \emu\ template for both the \mumu\ and \ee\ channels.

\begin{figure*}[!ht]
\unitlength=0.39\textwidth
\newcommand{\tagpic}[1]{\begin{picture}(00,00)\put(-0.8,0.6){\text{\bf (#1)}}\end{picture}}

\includegraphics[width=\the\unitlength]{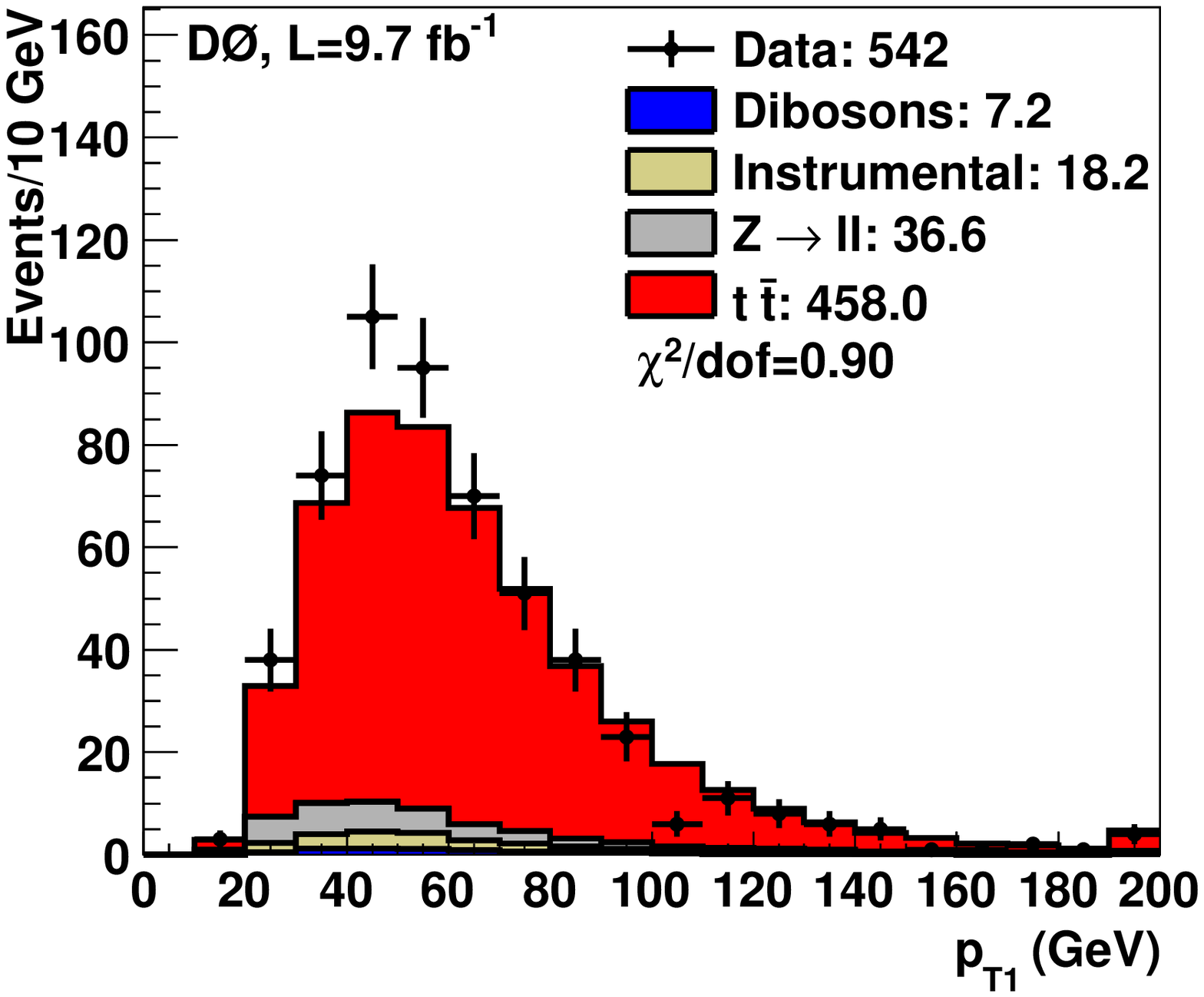}\tagpic{a}
\includegraphics[width=\the\unitlength]{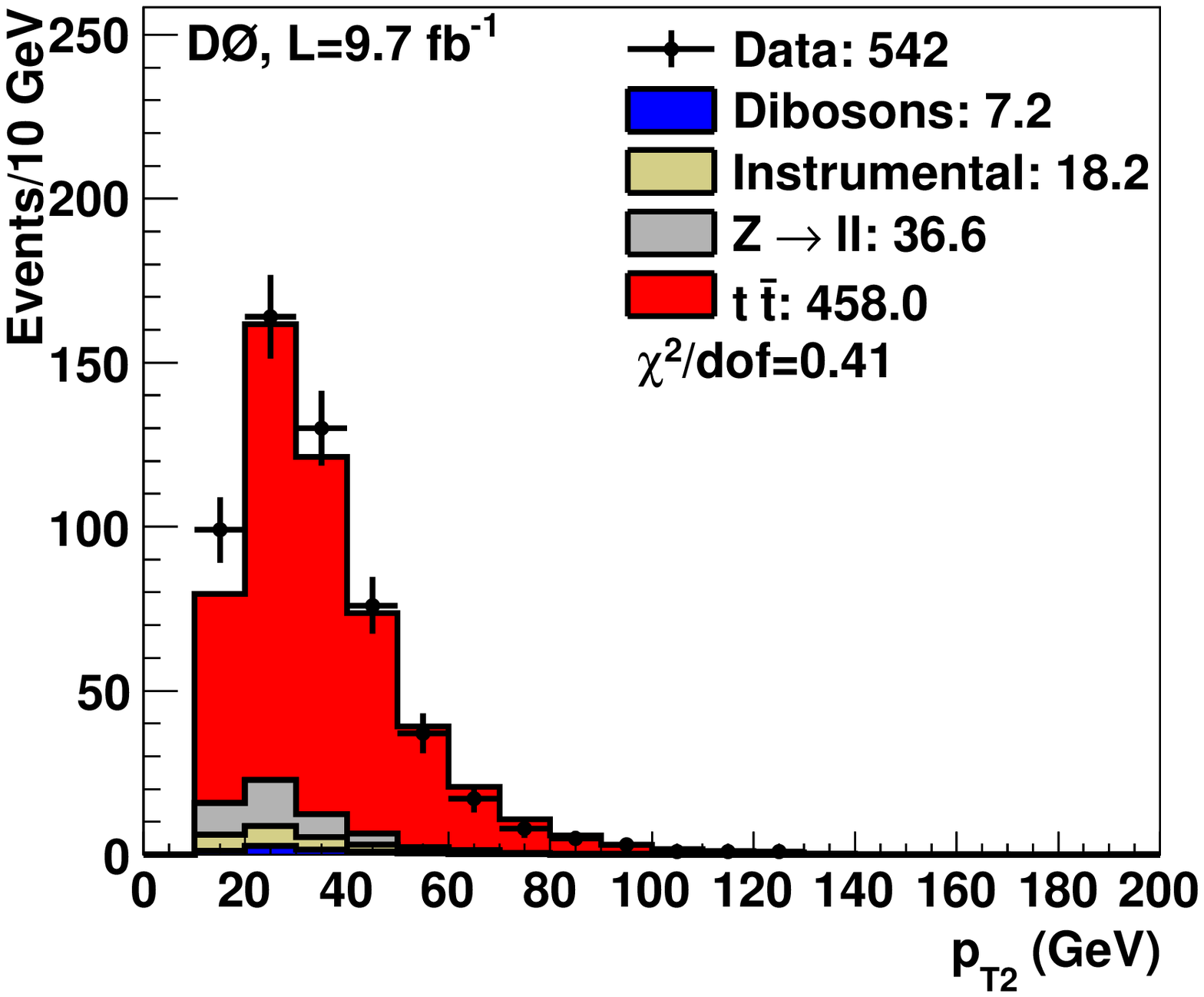}\tagpic{b}

\includegraphics[width=\the\unitlength]{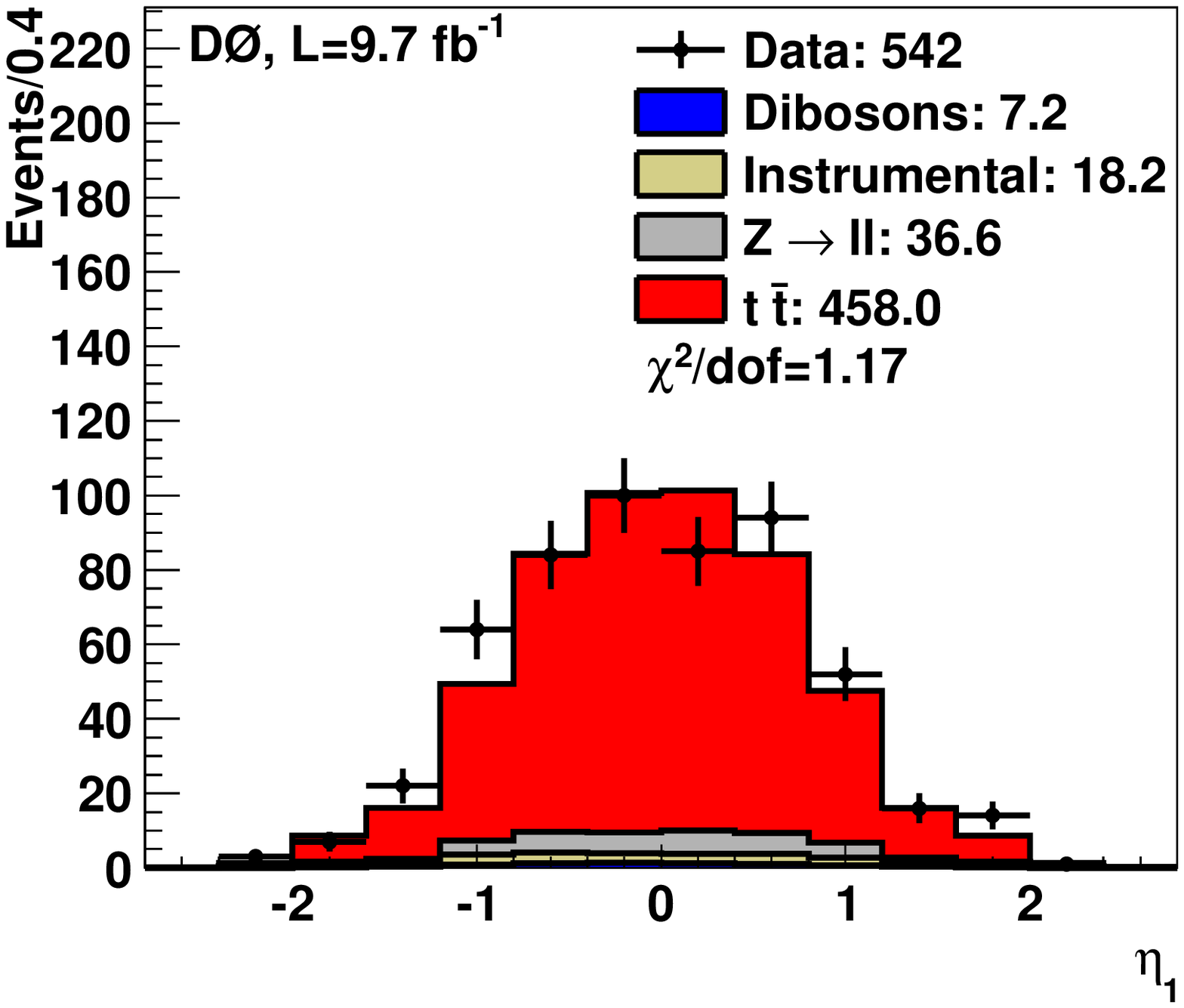}\tagpic{c}
\includegraphics[width=\the\unitlength]{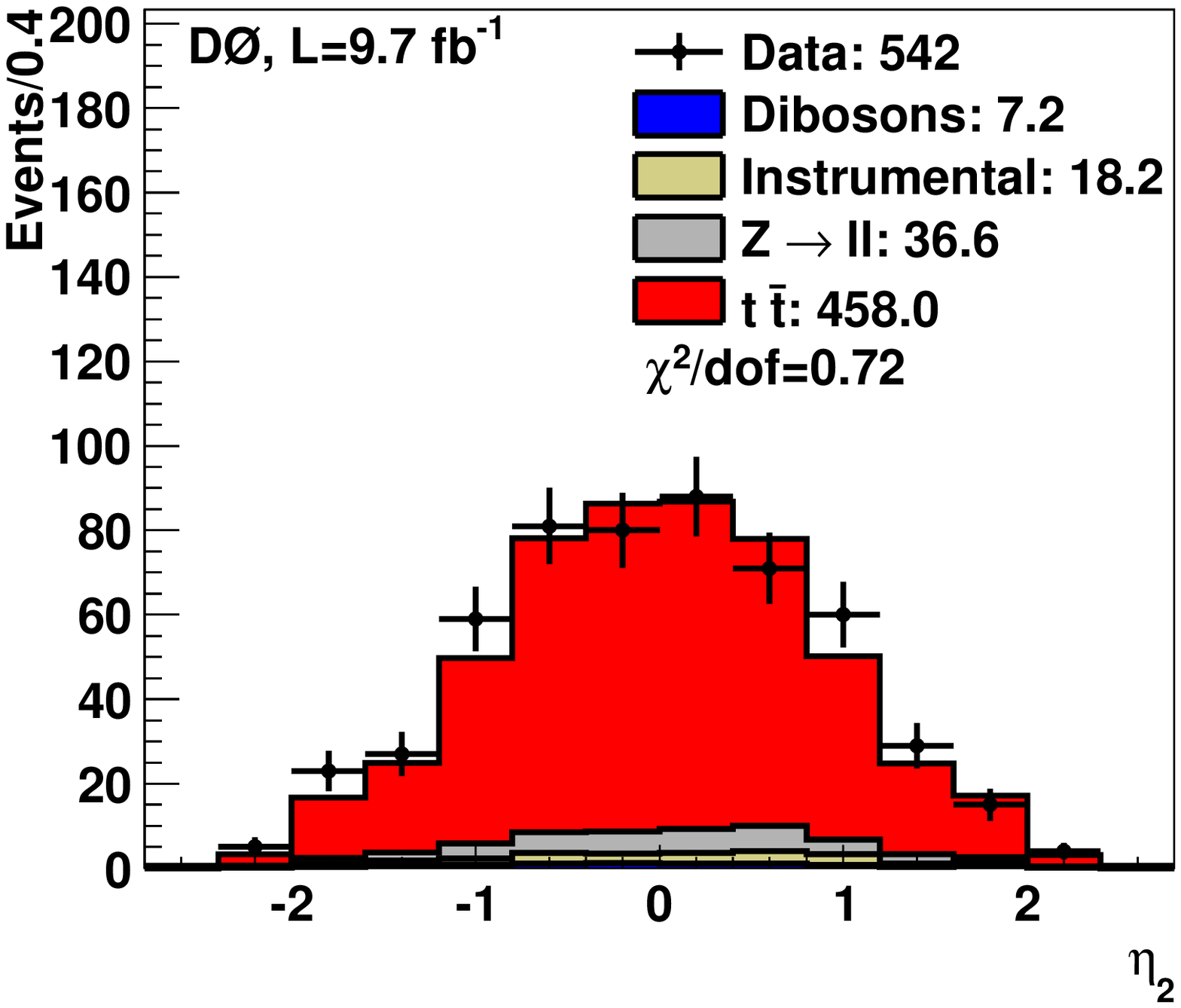}\tagpic{d}

\includegraphics[width=\the\unitlength]{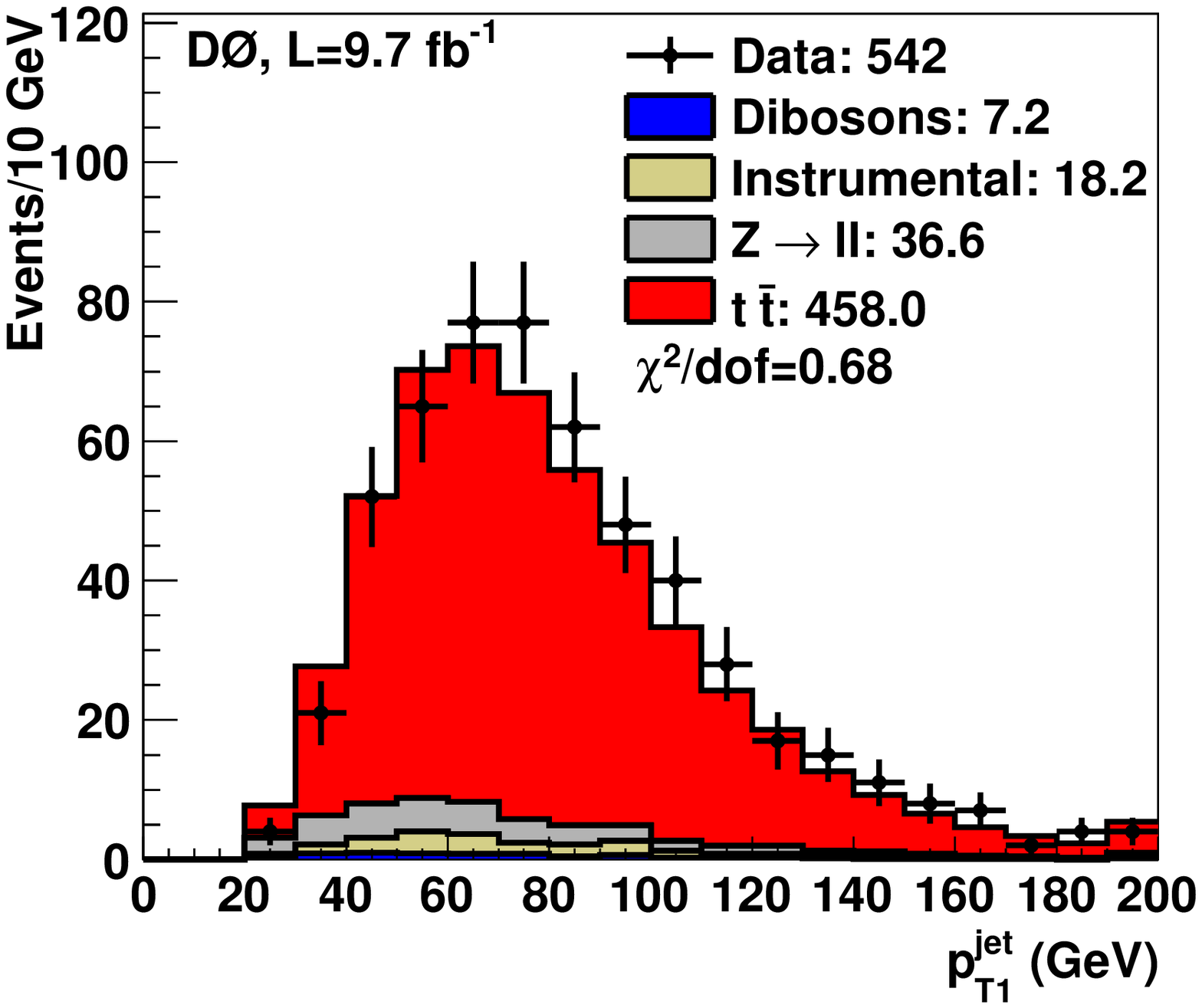}\tagpic{e}
\includegraphics[width=\the\unitlength]{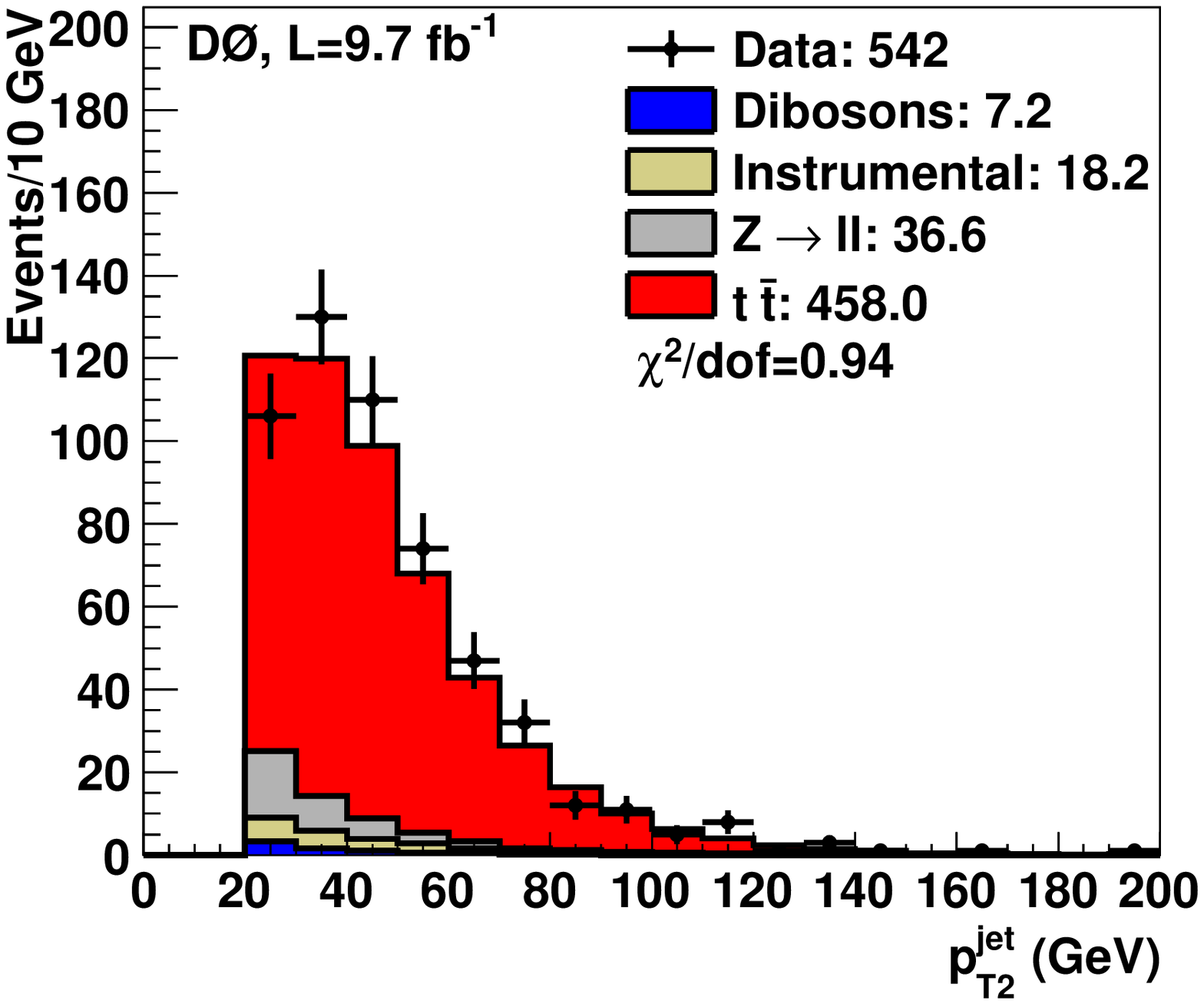}\tagpic{f}

\includegraphics[width=\the\unitlength]{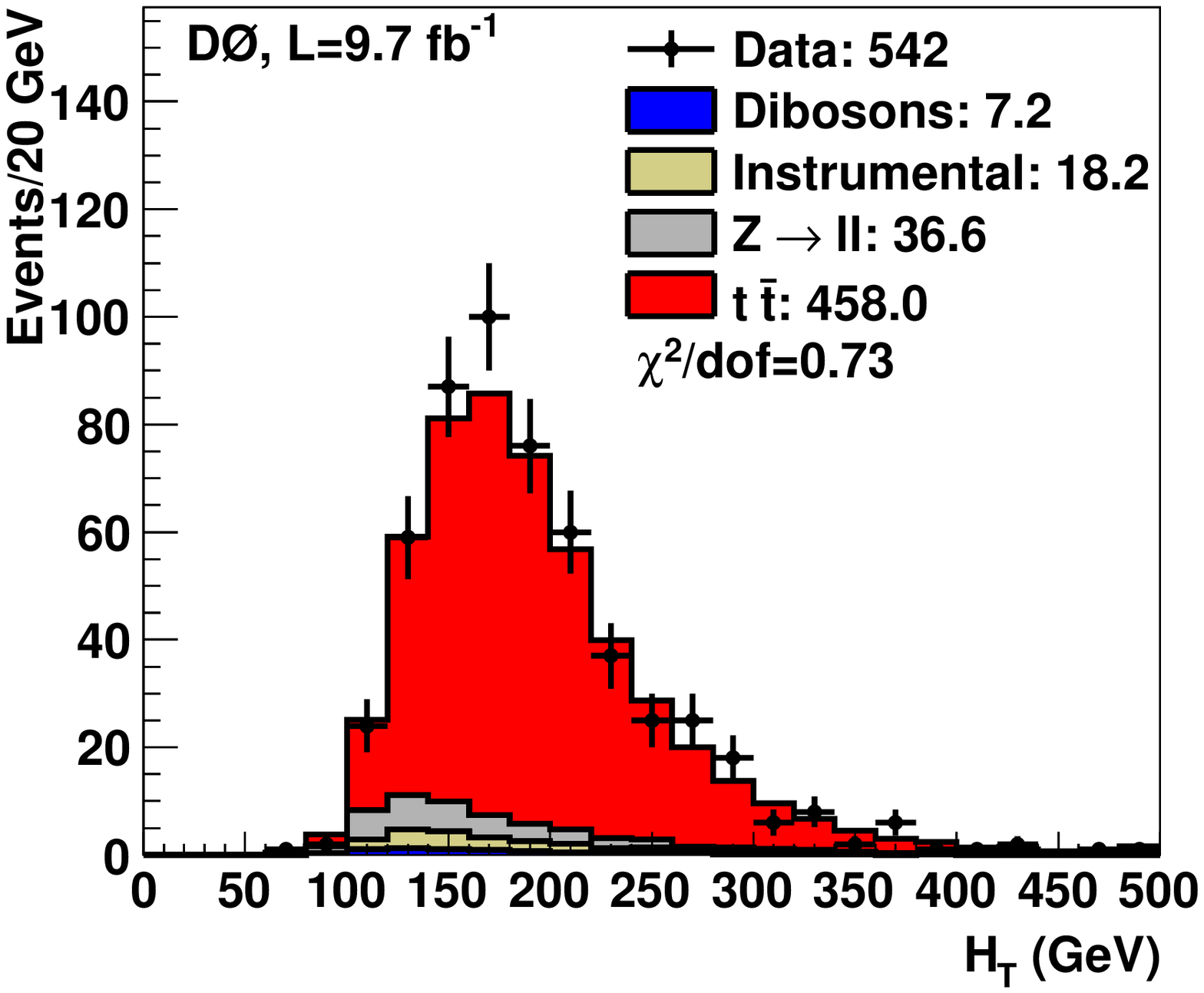}\tagpic{g}
\includegraphics[width=\the\unitlength]{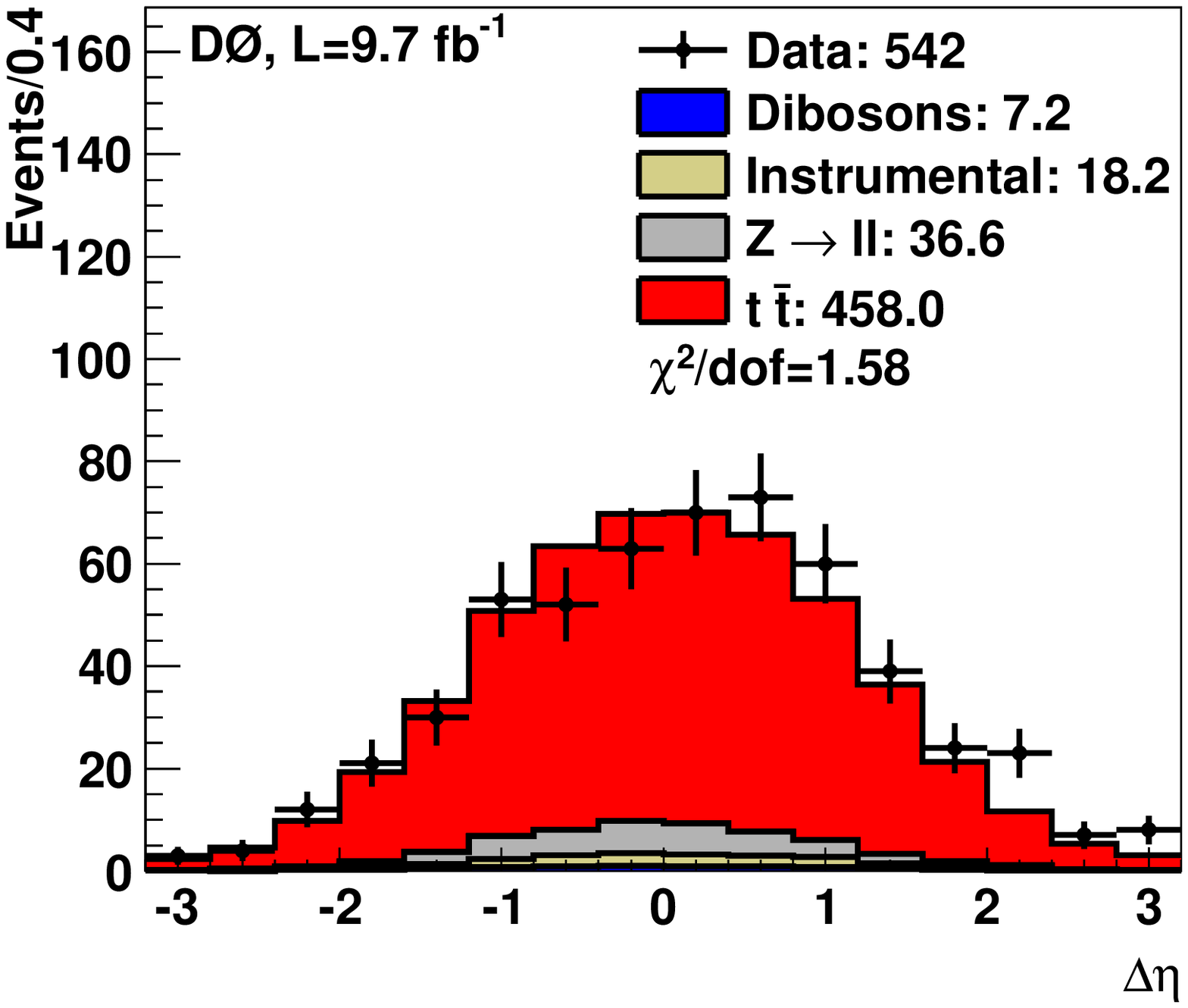}\tagpic{h}

 \caption{ [color online] Comparison of distributions between data  and MC simulations at the final selection for
(a) the transverse momentum of the leading lepton,
(b) the transverse momentum of the secondary lepton,
(c) the pseudorapidity of the leading lepton,
(d) the pseudorapidity of the secondary  lepton,
(e) the transverse momentum of the leading jet,
(f) the transverse momentum of the secondary jet,
(g) the $H_T$, and
(h) the difference between the two lepton pseudorapidities.
The overflow bin content has been added to the last bin.
\label{fig:selection_comb}}
\vspace{1cm}  \end{figure*}

\subsection{Comparison of MC simulation to selected data}
A comparison between  the expected and observed numbers of events 
at the final selection  levels is reported in Table~\ref{tab:cutflow}. 
The selected sample is relatively pure with a background fraction varying between 10\% and 16\% depending on  the channel.
\begin{table*}[!ht]
\caption{\label{tab:cutflow}
Comparison between expected and observed numbers of events at the final selection level for the different channels.  The  values are reported with
their statistical uncertainties.
}
\begin{tabular}{c c c c c c c c }
\hline\hline
Channel& \multicolumn{1}{c}{$Z\to \ell\ell$}  & \multicolumn{1}{c}{Dibosons}& \multicolumn{1}{c}{ Instrumental}
 & \multicolumn{1}{c}{$t\bar{t}\to \ell\ell jj$}
  &  \multicolumn{1}{c}{ Total expected }   &  \multicolumn{1}{c}{ Data  } &
\multicolumn{1}{c}{$\frac{\text{Data}}{\text{Expected}}$}
\\
\hline
 \mumu\  & $10.65^{+0.5}_{-0.5}$  & $1.7^{+0.1}_{-0.1}$  &    $\phantom{0}0.0^{+0.0}_{-0.0} $ &  $\phantom{0}79.3^{+0.6}_{-0.6}$  & $ \phantom{1}91.7^{+0.7}_{-0.7}$  & \phantom{1}92 & $ 1.00\pm0.10$
\\
 \emu\ & $13.03^{+0.5}_{-0.5}$  & $3.7^{+0.2}_{-0.2}$  & $16.4^{+0.7}_{-0.7} $ & $283.1^{+1.0}_{-1.0}$  &$ 316.2^{+1.3}_{-1.3}$  & $346$ & $1.09\pm0.05$\\ 
 \ee\ & $12.92^{+0.4}_{-0.4}$  & $1.9^{+0.1}_{-0.1}$  &$\phantom{0} 1.8^{+0.08}_{-0.08}$  & $ \phantom{1} 95.5^{+0.6}_{-0.6}$  & $112.1^{+0.8}_{-0.8} $ & 104 & $0.92\pm0.10$\\ 
\hline
\hline
\end{tabular}
\end{table*}
A comparison of kinematic distributions between data and expectations at the final selection level is shown in Fig.~\ref{fig:selection_comb}.
 
\makeatletter{}
\section{Matrix element method}
\label{sec:method}

\def\ddif{ {\rm d}}
\def\ddif{ {d}}

To reconstruct  distributions of kinematic observables describing the \ttbar\ events,
we use a novel modification of the matrix element (ME) integration 
developed for the $m_t$ measurements~\cite{Abazov:2011fc,Abazov:2015spa} by the \dzero\ Collaboration.
In particular, this method is employed to reconstruct the \dyttbar,
$\cos(\theta^+)$, and $\cos(\theta^-)$
 distributions, from which an estimate of the forward-backward asymmetry and top polarization are extracted.

\subsection{Matrix element integration}
The  ME integration  used in Refs.~\cite{Abazov:2011fc,Abazov:2015spa} consists in computing
the likelihood $L_z$ to observe a given event with the vector of measured quantities $z$, 
\begin{eqnarray}
L_z &= &\frac 1 { {\cal A}\cdot\sigma_{\mathrm{tot}}}  \sum_{\text{flavors}}\
\int\limits_{ \stackrel{x,q_{1},q_{2},}{  p^{\ttbar}_T ,\phi^{\ttbar}}}  
 f_ {{\rm PDF}}(q_{1})f_{{\rm PDF}}(q_{2})W(p^{\ttbar}_T) 
\nonumber\\ &\times&
 W(x,z)
\frac{(2 \pi)^{4} \left|\mathscr{M}\right|^{2}}
       {4 \sqrt{(q_{1} \cdot q_{2})^{2}}}
        {\ddif }\Phi_{6}
\ddif p^{\ttbar}_T\ddif\phi^{\ttbar} \ddif q_{1} \ddif q_{2}.
\label{eq:diff-probability}
\end{eqnarray}
In this expression, $x$ is  a vector describing the kinematic quantities
of the six  particles of the $p\bar p \to \ttbar \to \ell^+\nu b\,\ell^-\bar{\nu} \bar{b}$ final state,
 $\mathscr{M}$ is the matrix element describing the dynamics of the process,
 ${\ddif }\Phi_{6}$ is the 6-body phase space term,
the functions $f_{{\rm PDF}}$ are the PDFs of the incoming partons of  momenta $q_1$ and $q_2 $ and of different possible flavors,
$W(x,z)$, referred to as the {\it transfer function},
describes the probability density of a parton state $x$ to be reconstructed as $z$, 
$W(p^{\ttbar}_T)$ is a function describing the distribution of the $\ttbar$ system tranverse momentum, 
$p^{\ttbar}_T$, while the azimuthal angle of this system, $\phi^{\ttbar}$,
is assumed to have a uniform distribution over $[0,2\pi]$,
and  ${ {\cal A}\cdot\sigma_{\mathrm{tot}}}$ is the product of the experimental acceptance and the
production cross-section.
The matrix element,
$\mathscr{M}$, is computed  at leading order (LO) for $q\bar q$ annihilation only,
as it represents the main subprocess ($\approx 85\%$)  of the total $\ttbar$ production.
The functions   $f_{{\rm PDF}}$ are given by the {\textsc{CTEQ6L1}} leading order PDF set.
The function $W(p^{\ttbar}_T)$ is derived from parton-level simulated events generated with \alpgen\ interfaced to \pythia. More details on this function can be found in Ref.~\cite{Grohsjean:2008zz}.
Ambiguities between  partons and reconstructed particle assignments are properly treated by defining an effective transfer function that sums over all the different assignments
As we consider only the two leading jets in the integration process,
there  are only two possibilities to assign a  given jet to either the $b$ or $\bar b$ partons. 

The number of variables to integrate is  
given by the six three-vectors of final state partons (of known mass),
the  $\ttbar$ transverse momentum and transverse direction, and the longitudinal momenta of the two incoming partons.
These 22 integration variables are reduced by the following constraints:
the lepton and $b$-quark directions are assumed to be perfectly measured
(8 constraints),
the energy-momentum  between the initial state and  the final state is conserved (4 constraints), 
the $\ell^+\nu$ and $\ell^-\bar{\nu}$ system have a mass of  $M_W=80.4$~\gev~\cite{Agashe:2014kda} (2 constraints),
and the $\ell^+\nu b $ and $\ell^-\bar{\nu}\bar b$ system have a mass of  $m_t=172.5$~\gev\ (2 constraints).
Transfer functions account for muon and jet energies.
The transfer functions  are the same as used in Ref.~\cite{Abazov:2015spa}.
The electron momentum measurement has a precision of $\approx 3\%$,
which is much better than the muon momentum resolution of typically 10\% and the jet momentum resolution of  typically 20\%.
We thus consider that the electron momenta are perfectly measured.
This gives one additional constraint in the $\emu$ channel and two additional constraints in the \ee\ channel.
Thus, we integrate over 4, 5, and 6 variables in the \ee, \emu, and \mumu\ channels, respectively.
The integration variables are
 $ p^{\ttbar}_T$,
 $\phi^{\ttbar}$,
 energy of leading jet,
energy of sub-leading jet,
and  energy of the muon(s) (if applicable).

\label{sec:vegas_implementation}
The integration is performed using the MC-based 
numerical integration program \vegas~\cite{bib:VEGAS1,bib:VEGAS2}.
The interface to the \vegas\ integration algorithm is provided by the
GNU Scientific Library (GSL)~\cite{bib:GSL}. The MC integration consists of randomly sampling the space of integration variables, computing a weight for each of the random points that accounts for both
the integrand and the elementary volume of the sampling space, and finally summing all of the weights. 
The random sampling is based on a grid in the space of integration that is
iteratively  optimized to ensure fine sampling in regions with
large variations of the integrand.
For each of the random points, equations are solved to transform
these integration variables into the parton-level variables of Eq.~(\ref{eq:diff-probability}),   accounting for the measured quantities $z$.
The Jacobian of the transformation  is also computed to ensure proper weighting of the sampling space elementary volume.

\subsection{Likelihood of a parton-level observable}
\label{subsec:likelihoodME}

For any kinematic quantity $K$ reconstructed from the parton momenta $x$, for example $K(x)=\yt-\ytbar$,
we can build a probability density $L_z(K)$  that measures
the likelihood 
of $K(x)$ at the partonic level to give the  reconstructed  value $K$.
This likelihood is obtained by inserting a  term $\delta(K(x)-K)$
in the integrand of Eq.~(\ref{eq:diff-probability}), and normalizing the function so that $\int L_z(K) dK=1$.
The probability density is obtained by modifying
the \vegas\ integration algorithm.
For each reconstructed \ttbar\ event and 
each point in the integration space tested by \vegas, 
the integrand of 
Eq.~(\ref{eq:diff-probability}) and the quantity $K$ are computed.
After the full space of integration has been sampled, we obtain a weighted distribution of the variable $K$ that
represents the function  $L_z(K)$ up to an overall normalization factor.

\begin{figure*}[!ht]
\unitlength=0.33\textwidth
\newcommand{\tagpic}[1]{{\begin{picture}(00,00)\put(-0.75,0.6){\text{\bf (#1)}}\end{picture}}}
\includegraphics[width=\the\unitlength]{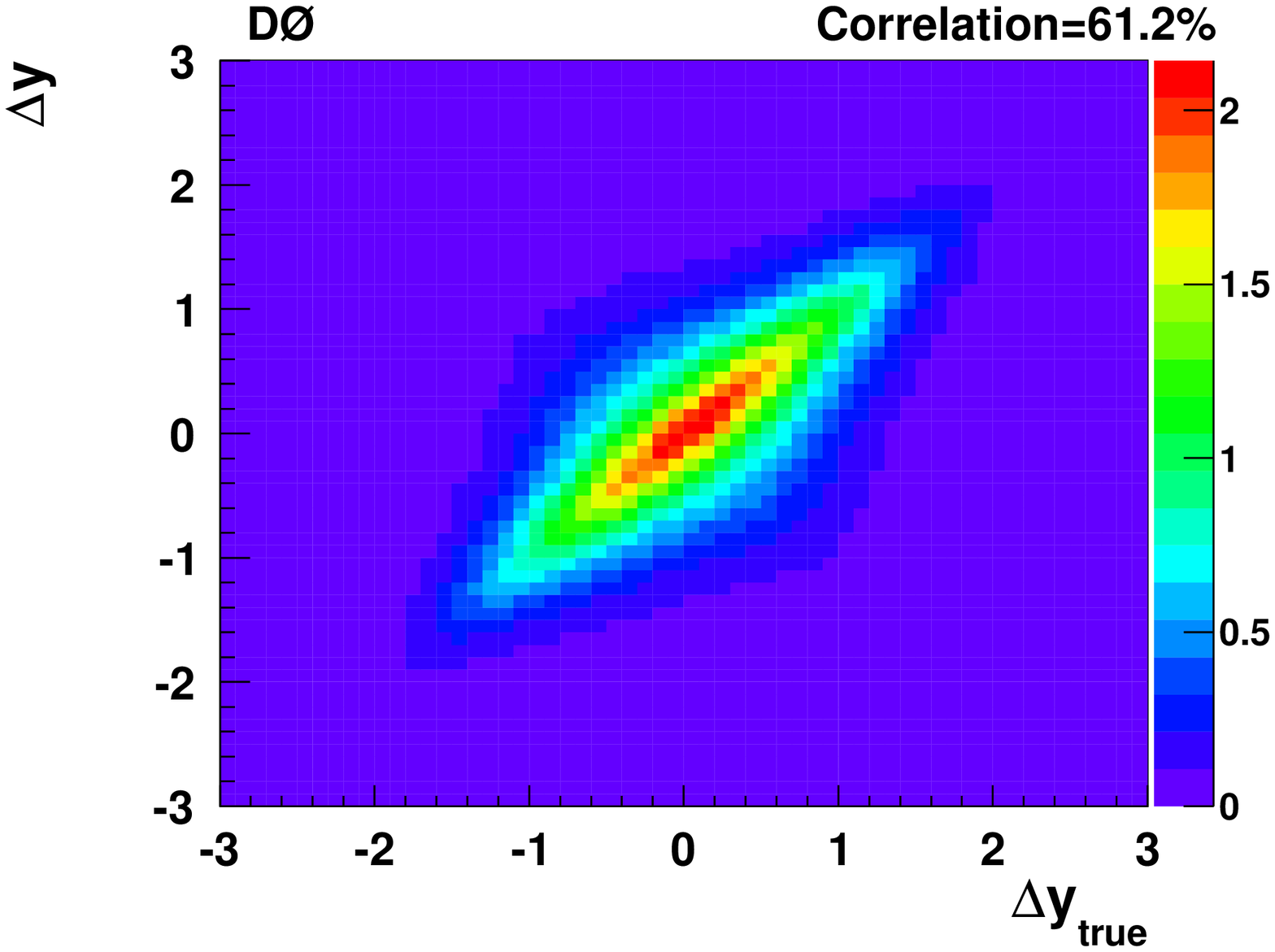}\tagpic{a}\includegraphics[width=\the\unitlength]{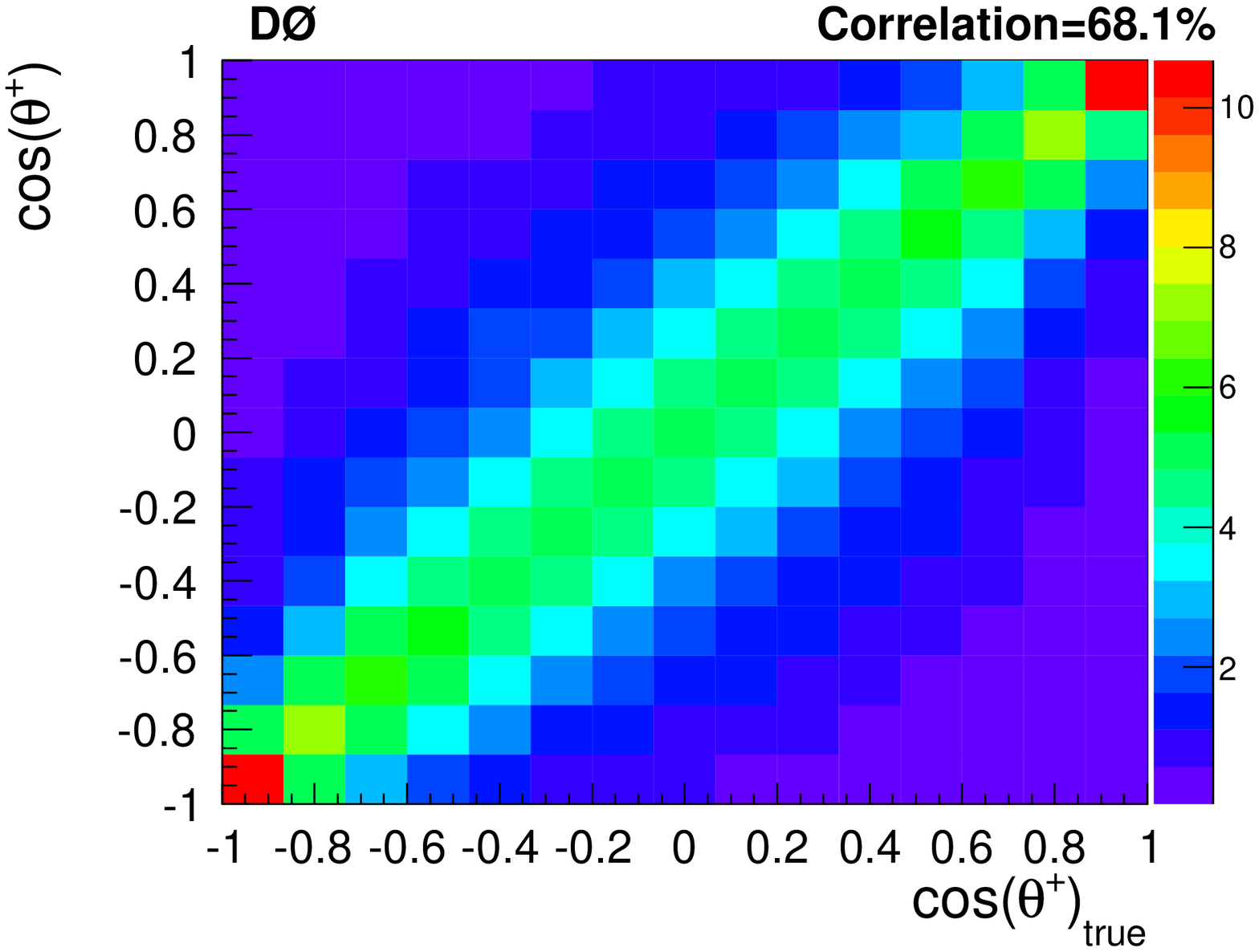}\tagpic{c}\includegraphics[width=\the\unitlength]{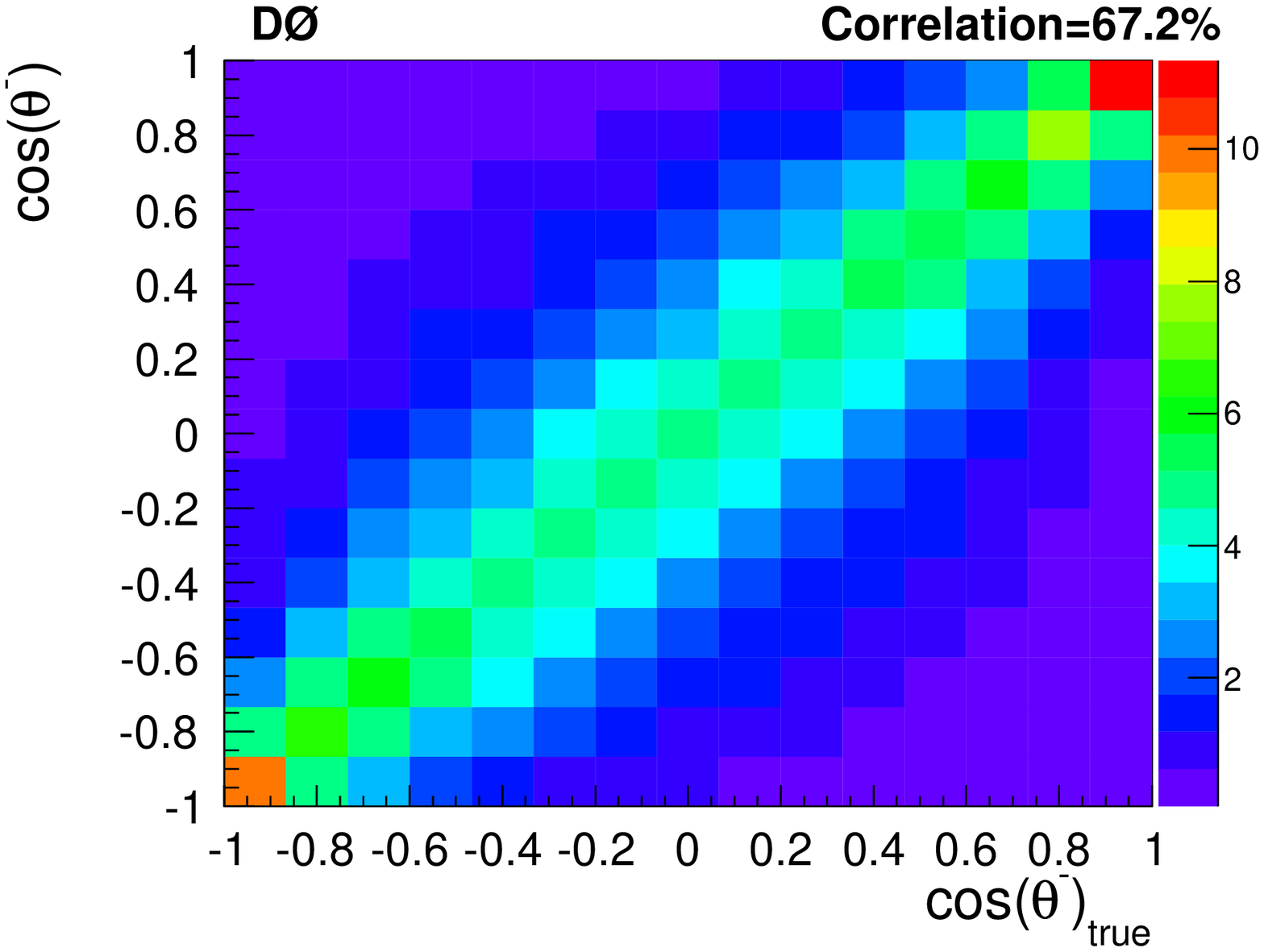}\tagpic{d}

\caption{\label{Fig:fig_2D_likelihood}
 [color online] Accumulation of likelihood functions ($\sum_{\text{events}} L_{z_i}(K)$, with $K$ along the vertical axis) versus the corresponding true parton level quantity ($K_{\mathrm{true}}$ along the horizontal axis) in \ttbar\ MC events after applying the selection criteria
for 
(a) $K=\dyttbar$, (b) $K=\cos\theta^+$, and (c) $K=\cos\theta^-$.
Each single MC event $i$ contributes in these plots with a complete distribution, $L_{z_i}(X)$, along the vertical axis for a given value on the horizontal axis, $K_{\mathrm{true}}$. The shades of color indicate the bin contents in arbitrary units.}
\end{figure*}

For each reconstructed event with observed kinematics $z_i$, where $i$ is an event index, we obtain a  likelihood function  $L_{z_i}(K)$.
By accumulating these likelihood functions over the sample of events,
we obtain a distribution that estimates the true distribution of the variable $K$.
The performance of this method  of reconstruction for parton-level distributions is estimated by comparing
the accumulation of likelihood functions 
to the true parton-level quantities for MC events, as shown 
in  Fig.~\ref{Fig:fig_2D_likelihood}.

\subsection{Raw estimate of \Att}
\label{sec:attraw_definition}
We could choose to use the maximum of the likelihood function $L_z(\dyttbar)$
to estimate the true value of $\dyttbar$ on an event-by-event basis.
However, to maximize the use of available information, we keep the full shape
 of the $L_z$ functions and accumulate these functions over the sample of \ttbar\ events to  obtain an estimate of the parton level distributions, which is then used to determine \Att. 
This method has been verified to perform better than the maximum likelihood method.
The distribution  $ \sum_{\text{events}} L_{z_i}(\dyttbar)$ is  shown
in Fig.~\ref{fig:sub_asymetry_COMB}(a), after subtracting the background contributions from the data.
The raw asymmetry \Attraw, extracted from this distribution, is reported in Table~\ref{tab:Asym_RAW}. Since this $\dyttbar$ distribution
is an approximate estimation  of the true distribution of $\dyttbar$,
 the raw asymmetry  $\Attraw$  is an approximation of the true \Att. The measurement therefore needs to be calibrated. The calibration is discussed below.

\begin{table}[!ht]
\caption{
\label{tab:Asym_RAW}
Raw forward-backward asymmetry  in  data before background subtraction, $\Adataraw $,
 asymmetry of the background,  $\Abkgraw$, and  measurement once the background contribution
has been subtracted,  $\Attraw$.
Asymmetries are reported in percent, together with their statistical uncertainties.
}

\renewcommand{\arraystretch}{1.2} 
\begin{tabular}{l + + +}
  \hline  \hline
  Channel       & \multicolumn{1}{c}{ \ \ \     \Adataraw }&\multicolumn{1}{c}{\ \ \ \Abkgraw} &\multicolumn{1}{c}{\ \ \ $\Attraw=\Adataminusbkgraw$}\\
   \hline
\emu  
&9.2 + 3.8  &0.3 + 1.9  &10.1 + 4.2  \\
\ee  
&15.8 + 6.4  &0.1 + 2.0  &18.8 + 7.6  \\
\mumu  
&6.7 + 7.9  &-0.3 + 3.3  &7.8 + 9.1  \\ \hline
Dilepton  
&10.1 + 3.0  &0.1 + 1.1  &11.3 + 3.4  \\
  \hline  \hline
\end{tabular}

\end{table}

The use of an event-by-event likelihood function
allows us to define an asymmetry observable for each event
\begin{equation}\label{eq:A_def}
A =  
{  \int_0^{\infty}L_{z} (\dyttbar) \ddif \dyttbar - \int^0_{-\infty}L_{z} (\dyttbar)} \ddif \dyttbar ,
\end{equation}
where the observable $A$ averaged over the sample of $\ttbar$ candidate events is
equal to the  raw asymmetry~\Attraw.
By construction,  $A$ lies in the interval $[-1,+1]$.
For a perfectly reconstructed event without resolution effects, $A$ would be either equal to $-1$  for $\dyttbar<0$ or to $ +1$ for $\dyttbar>0$.
The use
of $A$ allows us to determine
the  statistical uncertainty on  $\Attraw$ as the uncertainty on the average of a distribution.

\subsection{Raw estimate of  \Ptt}
\label{sec:Praw_definition}

\begin{figure*}[!ht]
\unitlength=0.35\textwidth
\newcommand{\tagpic}[1]{\begin{picture}(00,00)\put(-0.8,0.45){\text{\bf (#1)}}\end{picture}}
    \includegraphics[width=\the\unitlength]{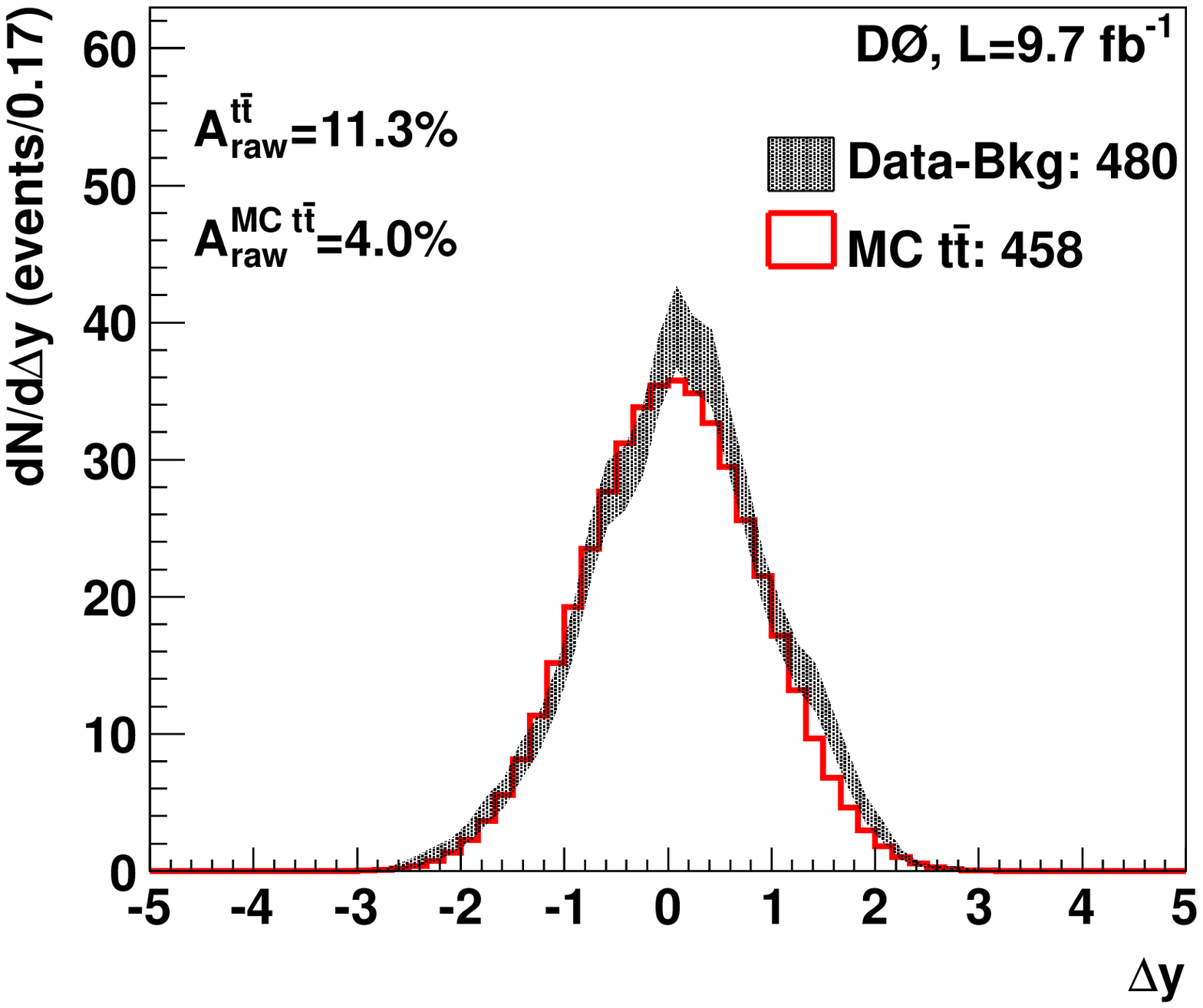}\tagpic{a}\includegraphics[width=\the\unitlength]{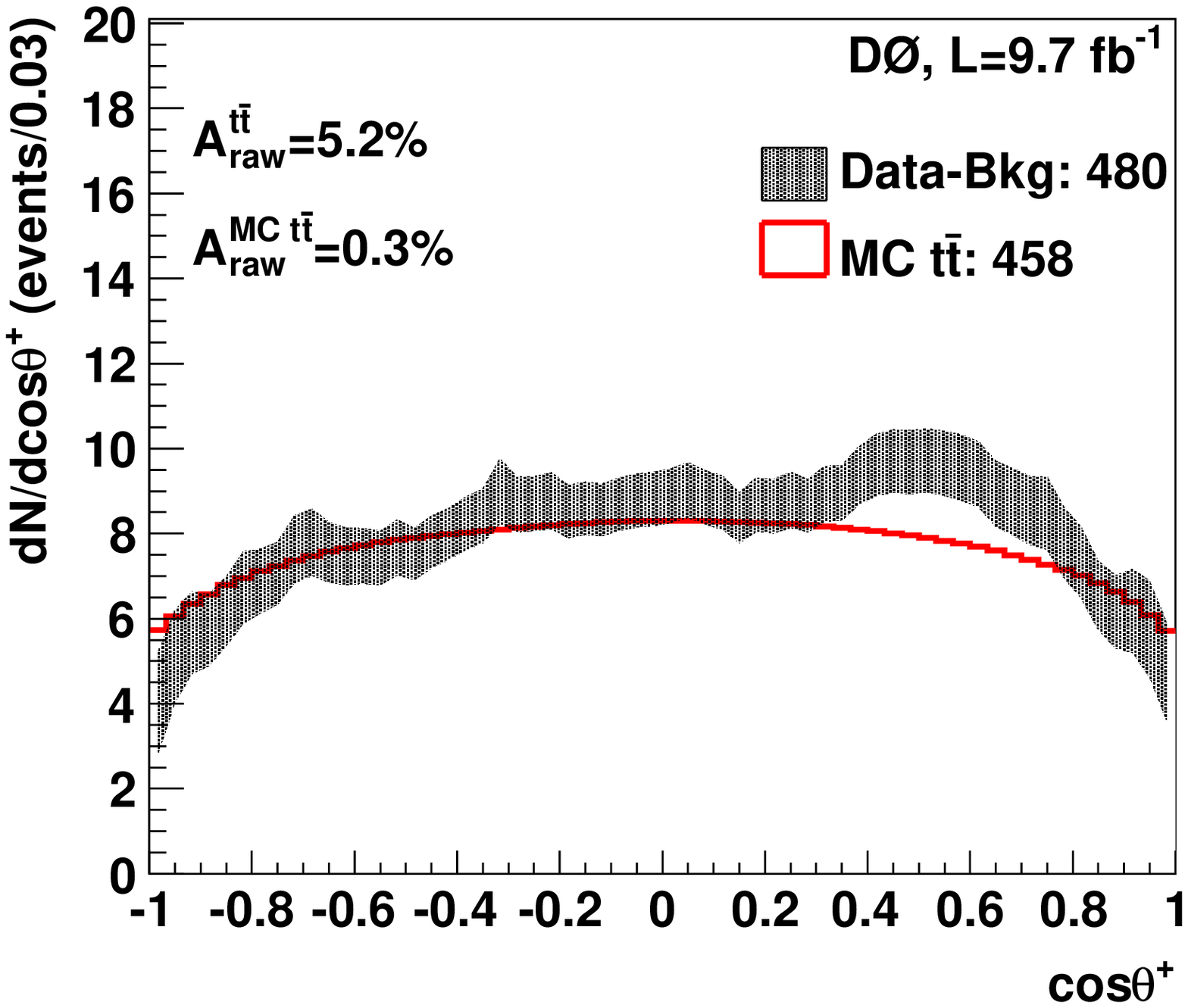}\tagpic{b}\includegraphics[width=\the\unitlength]{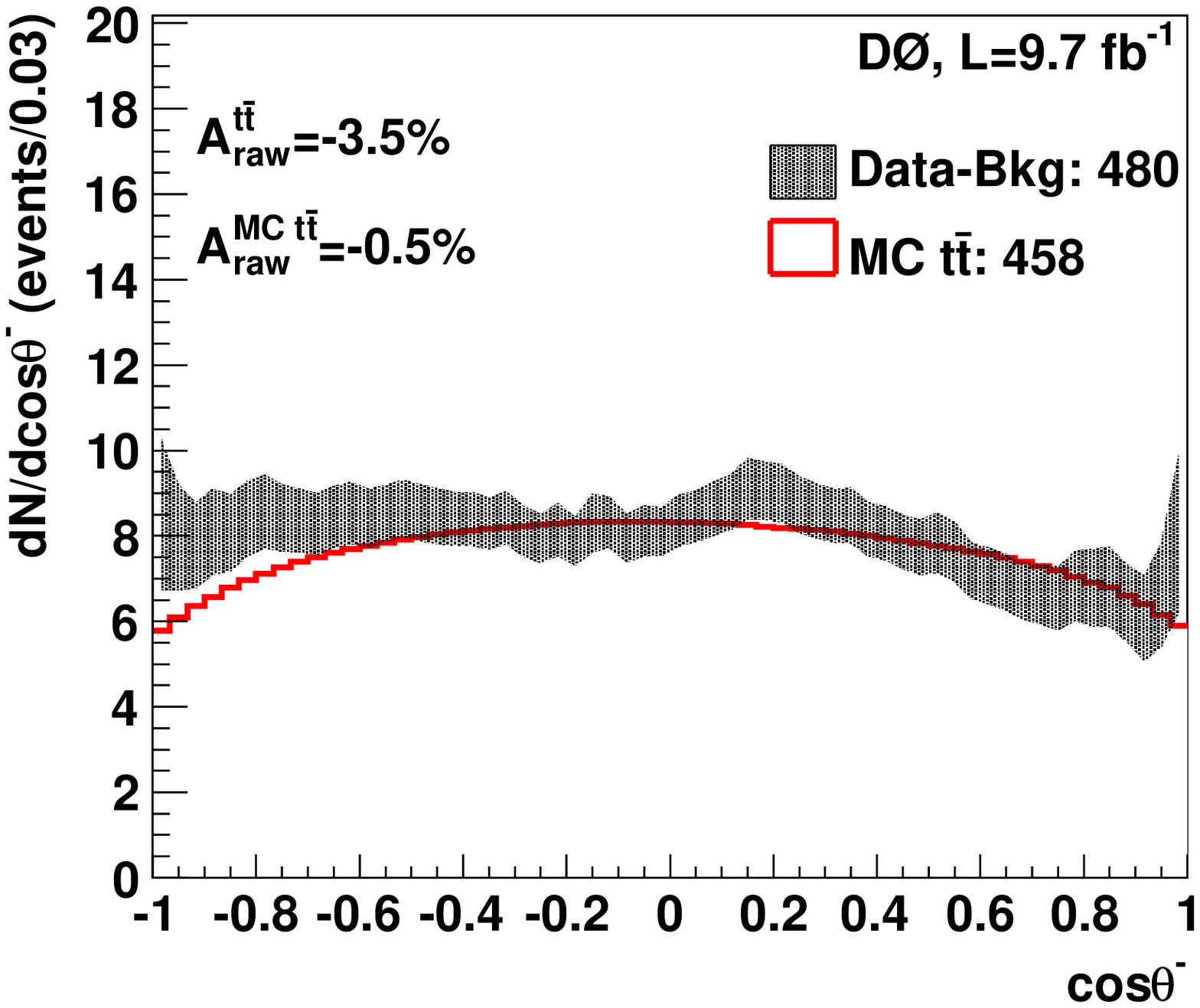}\tagpic{c}
 \caption{ [color online]
Estimated distribution  of the (a) \dyttbar, (b) $\cos\theta^+$, and  (c) $\cos\theta^-$ observables in dilepton events after subtracting the expected background contribution. Deviations beweeen the background-subtracted data and MC can be attributed to statistical fluctuations.
The background-subtracted data asymmetries and the MC asymmetries extracted from these distributions are also reported. These raw asymmetries need to be corrected for calibration effects to retrieve the parton-level asymmetries.
\label{fig:sub_asymetry_COMB}}
 \end{figure*}

In the same way as in the previous section, we use the accumulation  of the
likelihoods
$L_z(\cos\theta^+)$  and  $L_z(\cos\theta^-)$
to estimate the distributions of $\cos\theta^+$  and  $\cos\theta^-$.
The distributions  $\sum_{\text{events}} L_{z_i}(\cos\theta^{+})$  and    $\sum_{\text{events}} L_{z_i}(\cos\theta^{-})$ are shown in Figs.~\ref{fig:sub_asymetry_COMB}(b) and \ref{fig:sub_asymetry_COMB}(c), after subtracting the background contributions from the data.
The raw asymmetries, 
$\AP{+}{\invisible{\rm beam,} {\rm raw}}$
 and $\AP{-}{\invisible{\rm beam,} {\rm raw}}$,  and the raw polarization 
$ \Praw=\AP{+}{\invisible{\rm beam,} {\rm raw}}-\AP{-}{\invisible{\rm beam,} {\rm raw}}$ 
extracted from the data are reported in Table~\ref{tab:Polar_RAW}.
As for \Attraw, the measurement of \Praw\ needs to be calibrated to retrieve the 
parton-level values of the polarization.

\begin{table*}[!ht]
\caption{
\label{tab:Polar_RAW}
Asymmetry estimates
for the
$\cos\theta^{\pm}$
 distributions.
The raw asymmetry measurement in the data before background subtraction,
$\AAP{\pm }{,\,\rm data}{ \rm raw}$,
the asymmetry of the background,
\AAP{\pm}{,\rm\, bkg}{\rm  raw},
and the  measurement once the background contribution
has been subtracted, \AAP{ \pm}{\rm,\, data-bkg}{\rm raw},
are reported. The polarization estimates defined as
$\kappa\Polar{\rm xx }{\rm  raw}=\AAP{+ }{,\,\rm xx}{ \rm raw}-\AAP{- }{,\,\rm xx}{ \rm raw}$ are also given.
All values  are reported in percent, together with their statistical uncertainties.
}

\small
\renewcommand{\arraystretch}{1.4} \begin{tabular}{l+++++++++}
  \hline\hline

  Channel 	&\multicolumn{1}{c}{
\AAP{+}{\rm,\, data}{\rm raw}}&\multicolumn{1}{c}{\AAP{+}{\rm,\, bkg}{ \rm raw}}&\multicolumn{1}{c}{\AAP{+ }{\rm,\, data-bkg}{ \rm raw}}&\multicolumn{1}{c}{
\AAP{-}{\rm,\, data}{\rm raw}}&\multicolumn{1}{c}{\AAP{-}{\rm,\, bkg}{ \rm raw}}&\multicolumn{1}{c}{\AAP{- }{\rm,\, data-bkg}{\rm raw}}&\multicolumn{1}{c}{
$\kappa\Polar{ \rm data}{\rm  raw}$}&\multicolumn{1}{c}{$\kappa\Polar{\rm bkg }{\rm  raw}$}&\multicolumn{1}{c}{
{ $\Praw=$ $\kappa\Polar{{\rm data}-{\rm bkg} }{\rm  raw}$ }}\\
 \hline
\emu  
&5.7 + 4.1  &0.6 + 2.1  &6.2 + 4.6  &-3.3 + 4.1  &2.6 + 2.1  &-4.0 + 4.6  &9.0 + 5.8  &-2.0 + 2.4  &10.2 + 6.4  \\
\ee  
&13.4 + 7.2  &-3.2 + 2.0  &16.5 + 8.6  &-0.8 + 7.2  &-0.5 + 2.1  &-0.9 + 8.6  &14.2 + 10.1  &-2.7 + 2.3  &17.4 + 12.0  \\
\mumu  
&-9.4 + 8.1  &3.9 + 3.6  &-11.5 + 9.4  &-3.7 + 8.1  &2.3 + 3.5  &-4.7 + 9.3  &-5.7 + 11.8  &1.5 + 3.7  &-6.9 + 13.7  \\ \hline 
Dilepton  
&4.6 + 3.3  &0.2 + 1.3  &5.2 + 3.7  &-2.9 + 3.3  &1.7 + 1.2  &-3.5 + 3.7  &7.5 + 4.7  &-1.5 + 1.4  &8.7 + 5.3  
\\
  \hline \hline
\end{tabular}

\end{table*}

\subsection{Statistical correlation between \Attraw\ and  \Pttraw }
\label{sec:correlation_measurement}
We measure the statistical correlation between $\Attraw$ and  $\Pttraw$ in the data, which
is needed to determine the statistical correlation between
the measurements of \Att\ and \Ptt.
In the same way as $ \Attraw$ is the average of an event-by-event asymmetry $A$,
the raw asymmetries $\AP{+}{\invisible{\rm beam,} {\rm raw}}$ and $\AP{-}{\invisible{\rm beam,} {\rm raw}}$ 
are the averages  of event-by-event asymmetries denoted by $A_{\cos\theta^+}$ and $A_{\cos\theta^-}$.
The correlation between \Attraw\ and \Pttraw\ is identical to the correlation between the observables $A$ and $(A_{\cos\theta^+}-A_{\cos\theta^-})$.
This  correlation is determined  from the background subtracted data by computing the RMS and mean values of the distributions of $A$,  $(A_{\cos\theta^+}-A_{\cos\theta^-})$, and 
 $A\cdot (A_{\cos\theta^+}-A_{\cos\theta^-})$: 
\begin{align}
& {\rm cor}(\Attraw, \Pttraw)=& \nonumber \\&
\frac{  < A\cdot (A_{\cos\theta^+}-A_{\cos\theta^-})  > - \Attraw\cdot\Pttraw   }
{ {\rm RMS}(A) \cdot {\rm RMS}(A_{\cos\theta^+}-A_{\cos\theta^-}) }
.\end{align}
We report the values measured in  data in Table~\ref{tab:correlation_RAW}.

\begin{table}[!htb]

\caption{
\label{tab:correlation_RAW}
Measurement of the statistical correlation between the asymmetry 
 $\Attraw$ and the 
 polarization  \Praw
for the data, background, and background subtracted data.
Values  are reported in percent, together with 
 their statistical uncertainties.
}

\begin{tabular}{l+++}
  \hline \hline
  Channel 	&\multicolumn{1}{c}{ Data }&\multicolumn{1}{c}{ Background  }&\multicolumn{1}{c}{ Data$-$Background} \\
\hline
\emu  
&27 +\phantom{1} 6    & 9 + 3    &28 +\phantom{1} 6   \\% &25 + 0.4    \\
\ee  
&10 + 12    & 9 + 3    & 9 + 14 \\% &21 + 0.6    \\
\mumu  
&36 + 10    & 6 + 5    &39 + 12   \\% &34 + 0.6    \\
\hline
Dilepton  
&26 +\phantom{1} 5    & 9 + 2    &28 + \phantom{1}5    \\%&26 \pm 0.3    
  \hline \hline
\end{tabular}
\end{table}

\makeatletter{}
\section{Results corrected for calibration}
\label{sec:calibration}

The calibration procedure finds
a relation between the
raw asymmetry and polarization,  $(\Attraw,\Praw)$,
obtained after subtracting the background contributions,
and the true asymmetry and polarization $(\Attfb,\Ptt)$  of  \ttbar\ events.
The calibration procedure corrects for  dilution effects that arise from the limited  acceptance for \ttbar\ events, the finite resolution of the kinematic reconstruction, and
the simplified assumptions used in the matrix element integration (\eg, leading order ME, no $gg \to\ttbar$ ME, only two jets considered).
 The relation is inverted to extract a measurement of \Attfb\  and \Ptt\ from the values
of \Attraw\ and \Praw\ observed in  data.

The nominal calibration is determined using a sample of simulated \ttbar\ \mcatnlo\ dilepton events.
The procedure is repeated with
the samples from the other generators
(see section~\ref{sec:signal_generation} and~\ref{sec:BSM_description})
to determine different systematic uncertainties.
We  normalize the individual \ee, \emu, and \mumu\ contributions to have the same proportions as observed in the  data samples
after subtracting the expected  backgrounds.

\subsection{Samples for calibration}
\label{subsec:rewFunction}

We produce test samples from a nominal MC sample
by reweighting the events according to  the true value of the parton-level $\dyttbar$, $\cos\theta^+$, and  $\cos\theta^-$. The reweighting factors are computed as follows.

\subsubsection{Reweighting of lepton angular distributions}
\label{subsec:rewFunction_costheta}
The general expression for the  double differential lepton angle distribution
is~\cite{Bernreuther:2008ju}
\begin{eqnarray}
 \label{eq:two_lepton_polar}
\frac{\ddif^2 \sigma}
{\ddif \cos\theta^{+}\ddif \cos\theta^{-}}&=&
\frac 1 2 \left( 
1+ \kappa^{+} \Polar{+}{}\cos\theta^{+}\right. 
+ \kappa^{-} \Polar{-}{} \cos\theta^{-}\nonumber \\  &-& \left.C\cos\theta^{+}\cos\theta^{-}
\right),
\end{eqnarray}
where $C$ is the spin correlation coefficient, which is $\approx 90\% $ in the SM.
In the beam basis one has $\kappa\Polar{}{}\approx \kappa^{+} \Polar{+}{} \approx -\kappa^{-} \Polar{-}{}$. 
We use this relation
to reweight a given MC sample to simulate a target polarization of
 $\kappa\Polar{}{\rm  test}= \frac 1 2 \left( \kappa^{+} \Polar{+}{}{}-\kappa^{-} \Polar{-}{}{}\right)$.

\subsubsection{Reweighting of \dyttbar\ distribution}
\label{subsec:rewFunction_dy}
\begin{figure}[!ht]
\includegraphics[width=0.40\textwidth]{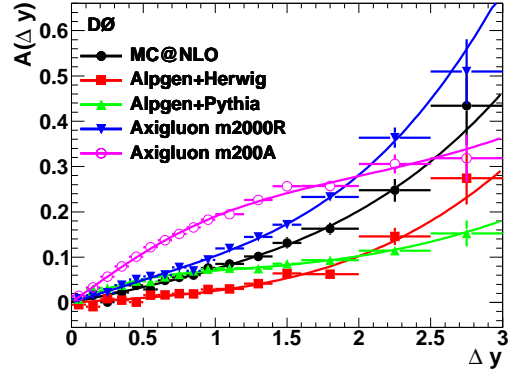}
 \caption{\label{fig:fit_DY_tanh+cubic} [color online]
 Differential asymmetry ${\cal A}(\dyttbar)$ at parton level for different MC samples. See Ref.~\cite{Carmona:2014gra} for the details on axigluon models.
 The observed  $ {\cal A}(\dyttbar)$ is fitted with the functional form of Eq.~(\ref{eq:tanh+cubic}).
}
\end{figure}

To determine a method of reweighting the \dyttbar\ distribution,
denoted ${\cal D}(\dyttbar) $,
we study
its shape using the different $\ttbar$ MC samples of
section~\ref{sec:signall_bsm_background}
at the generated level, \ie, before event selection and reconstruction.
Inspired by the studies performed for the distribution of rapidity of the charged leptons in  Ref.~\cite{AFB_form_discussion},
we rewrite ${\cal D}(\dyttbar) $ as
\begin{equation}
 {\cal D}(\dyttbar)=\frac 1 2 \left( {\cal D}(\dyttbar)+ {\cal D}(-\dyttbar)\right)\cdot \left(1+   {\cal A}(\dyttbar) \right),
\end{equation}
where ${\cal A}(\dyttbar)=\frac{ {\cal D}(\dyttbar)- {\cal D}(-\dyttbar) } { {\cal D}(\dyttbar)+ {\cal D}(-\dyttbar)}$
 is  the ratio between the  odd and even part of the $\dyttbar$ distribution, also called differential asymmetry as a function of
$\dyttbar$;
we then fit $ {\cal A}(\dyttbar)$  with 
an empirical odd function
\begin{eqnarray}\label{eq:tanh+cubic}
f(\dyttbar)
&=&
 \beta \times \tanh \left( \frac{\dyttbar}{\alpha} \right) +  \beta  \times \left(\frac{\dyttbar}{\gamma}\right)^3,
\end{eqnarray}
where $\alpha$ and $\gamma$ are shape parameters, while $\beta$ is a magnitude
parameter.  
The term $\beta  \times \left(\frac{\dyttbar}{\gamma}\right)^3$ was not needed in the study  of  Ref.~\cite{AFB_form_discussion}, but improves the modeling
significantly for the case of  $\dyttbar$.
The results of the fit for different \ttbar\ MC samples  are shown in Fig.~\ref{fig:fit_DY_tanh+cubic}.
If we reweight a MC sample so that the even part of the \dyttbar\ distribution, the term $\alpha$, and the term $\gamma$ are preserved,
then  the forward-backward asymmetry is proportional to $\beta$.

These considerations yield the following procedure to produce a sample
of test asymmetry $\Att_{\rm test}$ starting from a MC sample
of generated asymmetry  $\Att_{\rm sample}$.
We first fit the  differential asymmetry at the generated level
${\cal A}(\dyttbar)$ with the function $f$ of Eq.~(\ref{eq:tanh+cubic}) and determine
the parameters $\alpha$, $\beta$, and $\gamma$. Then we
apply weights to the events processed through  the \dzero\ simulation
\begin{align} \label{eq:reweighting_dy}
w(\dyttbar)= \frac
{  1+ \frac  {\Att_{\rm test}}{\Att_{\rm sample}} f(\dyttbar)}
{ 1+  f(\dyttbar)}
\end{align}
This procedure preserves
the even part of the distribution of \dyttbar.
It also preserves  the original shape of the differential asymmetry, but changes its magnitude to the desired value.

\subsubsection{Calibration}
\label{sec:calib_result}
Starting from the nominal \mcatnlo\ \ttbar\ sample,
we produce  test samples using  the product of the weights
defined in sections~\ref{subsec:rewFunction_costheta} and~\ref{subsec:rewFunction_dy}.
We use a grid of values for polarizations of $\kappa\Polar{}{}=(-0.2,-0.1,0,0.1,0.2)$
and asymmetries of $\Att=(-0.1,0,0.1,0.15,0.2,0.25)$ to obtain
30 samples
in addition to the unweighted nominal sample.
We apply  the
method of ME reconstruction 
to each of the 31 fully simulated samples and extract
a raw measurement  $(\Attraw,\Pttraw)$ associated with a given  parton-level  $(\Att,\Ptt)$.
A fit to the obtained set of points in the space $(\Att,\Ptt,\Attraw,\Pttraw)$ determines two affine functions that relate the
reconstructed quantities to the true quantities: $\Attraw=f_1(\Att,\Ptt)$ and $\Praw=f_2(\Att,\Ptt)$.
The affine functions fit  the 31 points well, with residuals $<0.1\%$.
We rewrite the affine relations using a matrix equation:
\begin{equation}\label{eq:Matrix_calibration}
\doublet{\Attraw}{\Praw}= C\cdot \doublet{\Att}{\Ptt} + O,
\end{equation}
where $C$ is a  $2\times2$   calibration matrix and $O$ is a vector of offset terms. The values of the matrix $C$ and $O$ are reported in Table~\ref{tab:calibSlope} for the  the different dilepton channels.
To determine the statistical uncertainties on
the calibration parameters,
we use an ensemble method. We split  the \mcatnlo\ samples into 100 independent ensembles and then repeat the calibration procedure for each of them.

\begin{table*}[!ht]\caption{Calibration parameters and their statistical uncertainties for the different channels.
}
\label{tab:calibSlope}
\begin{tabular}{lcc}
\hline
\hline
  Channel 		& Calibration matrix $C$ 	& Offset $O$ \\
  \hline
  \emu		 & $ \matrixfour{0.617  \pm 0.008   }{ 0.148  \pm 0.002  }{ 0.346  \pm 0.008  }{0.560  \pm 0.003   }  $  & $  \doublet{\phantom{-}0.011  \pm 0.002   }{ -0.009  \pm 0.003  } $  \\
 \hline
  \ee 		        & $ \matrixfour{0.599  \pm 0.006   }{ 0.135  \pm 0.003  }{ 0.315  \pm 0.007  }{0.544  \pm 0.005   }  $  & $  \doublet{\phantom{-}0.007  \pm 0.003   }{ -0.003  \pm 0.005  } $  \\
  \hline
  \mumu 	& $ \matrixfour{0.639  \pm 0.007   }{ 0.189  \pm 0.004  }{ 0.460  \pm 0.008  }{0.649  \pm 0.006   }  $  & $  \doublet{\phantom{-}0.006  \pm 0.005   }{ -0.006  \pm 0.007  } $  \\ 
\hline
  Dilepton 	 & $ \matrixfour{0.617  \pm 0.008   }{ 0.153  \pm 0.002  }{ 0.359  \pm 0.006  }{0.572  \pm 0.002   }  $  & $  \doublet{\phantom{-}0.010  \pm 0.002   }{ -0.007  \pm 0.003  } $  \\  
  \hline  \hline
\end{tabular}

\end{table*}

\label{sec:ensemble_calib}

\subsection{Measurement of \Attfb\ and \Ptt\ after calibration}
\label{sec:att_measurement}

The calibration relation of Eq.~(\ref{eq:Matrix_calibration}) is inverted
to retrieve the true partonic asymmetry $\Attfb$ and the true polarization \Ptt\ from the reconstructed  $\Attraw$ and \Praw.
We obtain 
a measurement of $(\Attfb,\Ptt)$ reported in Table~\ref{tab:finalAsym}  for each dileptonic channel using the calibration coefficients from 
Table~\ref{tab:calibSlope} and the raw measurements from  Tables~\ref{tab:Asym_RAW} and~\ref{tab:Polar_RAW}.

\label{sec:top_mass_correction}
Two \alpgen+\pythia\ \ttbar\ samples generated at different $m_t$ are used to estimate the dependence of the measurement on $m_t$.
Considering a  top mass of $m_t=173.18\pm0.94~\gev$~\cite{Aaltonen:2012ra} as reference,
the dilepton results reported in Table~\ref{tab:finalAsym} have to be corrected by
$ -0.02\%$ and $0.15\%$ for  \Att\ and \Ptt, respectively.
The corrected combined dilepton results are
\begin{equation}
\Att=(15.0 \pm 6.4 \text{ (stat)})\%
\end{equation}
\begin{equation}
\Ptt=(7.2 \pm 10.5 \text{ (stat)})\%.
\end{equation}

\begin{table*}[!ht]\caption{
\label{tab:finalAsym} Measurements of $\Attfb$ and \Ptt\  for each dileptonic channel corrected for the calibration (for $m_t= 172.5~\gev$).
The statistical correlation between the two measurements arises both from the statistical
correlation of the experimental observables and the correction for the calibration.
}
\begin{tabular}{l+++}

  \hline
  \hline
  Channel 		&\multicolumn{1}{c}{ \parbox{0.2\textwidth}{$\Attfb$ (\%)} }&\multicolumn{1}{c}{ \parbox{0.15\textwidth}{$\Ptt$ (\%) } }&\multicolumn{1}{c}{   \parbox{0.2\textwidth}{statistical\phantom{(\%)}\\ correlation (\%)}}\\
  \hline
\emu			& 11.6 + \phantom{1}7.8 (\text{stat}) 
	                & 12.6 + 13.0 (\text{stat}) 
                        &  -48  \\
  \hline

\ee	
&26.1 + 15.2 (\text{stat}) 
&17.5 + 26.0 (\text{stat}) 
&-58   \\

  \hline
 \mumu
&17.8 + 16.7 (\text{stat}) 
&-22.2 + 24.6 (\text{stat}) 
&-52 \\

   \hline 
Dilepton
&15.0 +\phantom{1} 6.4 (\text{stat}) 
&7.0 + 10.5 (\text{stat}) 
&-50\\

  \hline
  \hline
\end{tabular}

\end{table*}

\makeatletter{}
\section{Systematic uncertainties}
\label{sec:systematics}

We consider three categories of uncertainties.
Uncertainties affecting the signal
are obtained by deriving  calibration coefficients from  alternate signal models and propagating
them  to the final results.
Uncertainties affecting the background  have an impact on the raw measurements,
\Attraw\ and \Pttraw, as these observables are obtained after subtracting the background.
They
are propagated to the final measurement by applying the nominal calibration correction to the modified \Attraw\ and \Pttraw.
The third category consists of the uncertainties on the calibration method.
Since the measurement is performed after background subtraction,
the calibration is independent of the normalization of the \ttbar\ simulation,
and there is no systematic uncertainty due to signal normalization.
The uncertainties on $\Attfb$ and \Ptt\ due to the different sources  are summarized in Table~\ref{tab:systematics}, together with the  correlations.

\subsection{Uncertainties on signal}
Several sources of systematic uncertainties due to the detector and reconstruction model affect the jets and thus the signal kinematics.
We consider uncertainties on the jet energy scale, flavor-dependent jet response, and jet energy resolution~\cite{Abazov:2013hda}.
We also take into account uncertainties associated with  $b$ tagging  and vertexing~\cite{NIM_btaging}.

To estimate the impact of higher order correction,
we compare  the calibration obtained with  \mcatnlo+\herwig\ to the calibration obtained with \alpgen+\herwig.
To propagate uncertainty on the simulation of initial state and final state radiations (ISR/FSR), the 
amount of radiation is varied  by  scaling the \texttt{ktfac} parameter either by a factor of $1.5$ or  $1/1.5$ in an \alpgen+\pythia\ simulation of \ttbar\ events~\cite{Abazov:2015spa}.
The hadronization and parton-shower model uncertainty
is derived from the difference between the \pythia\ and \herwig\ generators,
estimated by comparing \alpgen+\herwig\ to \alpgen+\pythia\ \ttbar\ samples.
The different models for parton showers used by various  MC generators yield
different amounts of ISR between forward and backward events~\cite{Skands:2012mm,Winter:2013xka}.
The uncertainty on the ISR model is defined
as  50\% of the difference between the nominal results and 
the results derived from a \mcatnlo\ simulation
in which the dependence of the forward-backward asymmetry on the $\pt$ of the $\ttbar$ system is removed.
The uncertainty of 0.94 GeV on $m_t$~\cite{Aaltonen:2012ra}  is propagated to the final result using two \alpgen+\pythia\  samples generated with different $m_t$ values.
We determine PDF uncertainties by varying the 20 parameters describing the \textsc{CTEQ6M1} PDF~\cite{Nadolsky:2008zw} within their uncertainties.

\subsection{Uncertainties on  background}

The uncertainty on the background level is  obtained by varying  the instrumental background normalization by 50\%
and
the overall background normalization by 20\%.
The model of the instrumental background kinematics  is varied, using the same method as in Ref.~\cite{Abazov:2013wxa}.
We  reweight  the reconstructed $\Delta y$, $\cos\theta^+$, and $\cos\theta^-$ distributions by  a factor of $1+\epsilon\times \sigma_{\rm band}$,
where $ \sigma_{\rm band}$ is the statistical uncertainty band of the distribution and $\epsilon=\pm 1$ is chosen to be
positive  for $\Delta y>0$, $\cos\theta^+>0$, $\cos\theta^-<0$, and negative for $\Delta y<0$, $\cos\theta^+<0$, and $\cos\theta^->0$.

\subsection{Uncertainties on calibration}
\label{sec:syst_functionnal_form}
We also consider sources of uncertainties affecting the calibration procedure.
The statistical uncertainty on the calibration parameters and their correlations  are propagated to the final measurements. The uncertainties are 0.60\% 
for \Att\ and 0.61\% for \Ptt. The correlation is $-39\%$.

To estimate a systematic uncertainty due to the choice of  \dyttbar\ calibration procedure, we reweight the \mcatnlo\ sample
to reproduce the shape of the  differential asymmetries of each different  BSM and SM model considered.
Each of the resulting samples serves as a seed for a new calibration procedure as described in  section~\ref{subsec:rewFunction_dy}.
The maximum variation  in the  \Att\ measurement obtained with these new calibrations is taken as systematic uncertainty.
It is obtained using the shape from the $\alpgen+\pythia$ sample and
amounts to 1.3\%. The impact of these tests is negligible for \Ptt\
since only the \dyttbar\ distribution is modified.

We also perform a closure test using the five different BSM models described in section~\ref{sec:BSM_description}. For each of the considered BSM models  we create  test samples by reweighting the \dyttbar\ and $ \cos\theta^{\pm}$ distributions, in the same way as described in section~\ref{subsec:rewFunction} for \mcatnlo\ samples. The samples cover a range of  values of  \Att\ and \Ptt\  centered around the data measurement within  $\pm 1$  statistical standard deviations.
These samples are treated as pseudo-data: We compute the
differences between  what would be measured using the nominal calibration  and the true \Att\ and \Ptt\ of each sample.
The maximum  \Att\ bias is found for the axigluon $m200L$ sample~\cite{Carmona:2014gra} and corresponds to a shift of $(\Delta \Att,\Delta \Ptt)= ( -2.9\%, 2.3\%)$ obtained for $(\Att,\Ptt)\approx (19\%,9\%)$. 
The maximum  \Ptt\ bias is found for the axigluon $m200A$ sample~\cite{Carmona:2014gra} and corresponds to $(\Delta \Att,\Delta \Ptt)= ( -1.5\%, 2.6\%)$  for $(\Att,\Ptt)\approx (10\%,0\%)$.
These two doublets in $(\Delta \Att,\Delta \Ptt)$ are taken as uncorrelated  systematic uncertainties. In each of these doublets, the uncertainty on  $\Att$ and $\Ptt$ are taken as $-100\%$ correlated.

\begin{table*}[!ht]
\caption{\label{tab:systematics} Summary of systematic and statistical uncertainties.}
\begin{tabular}{lccc}
  \hline \hline
Source of uncertainty			& Uncertainty on \Att\ (\%) & Uncertainty on \Ptt\ (\%)  & Correlation  (\%)  \\
\hline  \em{Detector modeling}           & & & \\
\ Jet energy scale             &   0.13  & 0.50   & $-100$ \\ 
\  Jet energy resolution       &   0.03  & 0.06   & $\phantom{-}100$ \\ 
\  Flavor-dependent jet response &   0.02  & 0.06   & $-100$ \\ 
\  $b$ tagging                 &   0.14  & 0.43   &  $-\phantom{0}94$ \\ 

\hline  \em{Signal modeling}             & & & \\
\ ISR/FSR                      &   0.16  & 0.41   & $-100$ \\ 
\ Forward/backward ISR         &   0.10  & 0.07   & $-100$ \\ 
\ Hadronization and showering  &   3.28  & 1.94   & $-100$ \\ 
\ Higher order correction      &   0.02  & 0.71   & $-100$ \\ 
\ PDF                          &   0.12  & 0.30   & $-\phantom{0}98$ \\ 
\ Top quark mass               &   0.03  & 0.21   & $-100$ \\ 
\hline  \em{Background model}            & & & \\
\ Instrumental background shape        &   0.16  & 0.53   & $\phantom{-}100$ \\ 
\ Instrumental background normalization &   0.29  & 0.01   &  $\phantom{-}100$ \\ 
\ Background normalization     &   0.44  & 0.18   &  $\phantom{-}100$ \\ 
\hline  \em{Calibration}                 & & & \\
\ \dyttbar\ model               &   1.28  & 0.11   &  $\phantom{-}100$ \\ 
\ MC statistics                &   0.60  & 0.61   & $-\phantom{0}39$ \\ 
\hline  \em{Model dependence}            & & & \\
\ Maximum \Att\ variation          &   2.91  & 2.35   & $-100$ \\ 
\ Maximum \Ptt\ variation          &   1.49  & 2.58   & $-100$ \\ 
\hline \em{Statistical uncertainty} &   6.40  & 10.53   & $-\phantom{0}50$ \\ 
    \hline 
{Total    systematic without model dependence} &   3.62  & 2.40   & $-\phantom{0}74$ \\ 
 \hline 
{Total    systematic} &   4.88  & 4.24   & $-\phantom{0}83$ \\ 
 \hline 
{Total    }       &   8.05  & 11.35   & $-\phantom{0}56$ \\ 
 \hline  \hline
\end{tabular}
\end{table*}

\makeatletter{}
\section{Results}\label{sec:measresults}\label{sec:conclusion}

\begin{figure}[!htb]
\center

\includegraphics[width=0.45\textwidth]{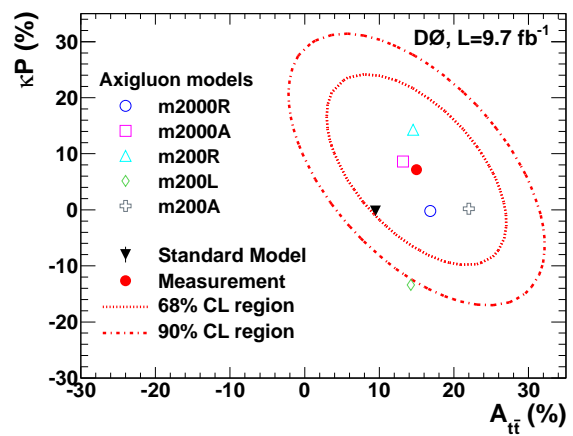}

\caption{\label{fig:2d_result_afb_polar} [color online]
Two dimensional visualization of the \Att\ and \Ptt\ measurements and comparison with benchmark axigluon models~\cite{Carmona:2014gra}.}
\end{figure}

The measurements and the uncertainties discussed in the previous sections are summarized by
\begin{eqnarray}
\Att &=& (15.0 \pm 6.4 \text{ (stat)} \pm 4.9 \text{ (syst)})\%,\nonumber\\
\Ptt &=& (7.2  \pm 10.5 \text{ (stat)} \pm 4.2 \text{ (syst)})\%,   \label{eq:final_result}
\end{eqnarray}
with a correlation of $-56\%$ between the  measurements.
The results are presented in Fig.~\ref{fig:2d_result_afb_polar}.
The NLO SM prediction for \Attfb\ is $\Att =  (9.5\pm 0.7)\% $~\cite{Czakon:2014xsa}, while  the SM polarization is expected to be small,  $\Ptt=(-0.19\pm 0.05)\% $~\cite{Bernreuther:2006vg}. Our measurement is consistent with the SM prediction within the 68\% confidence level region. 
In Fig.~\ref{fig:2d_result_afb_polar} we overlay the expected values for the different axigluon models of  Ref.~\cite{Carmona:2014gra}.
As the models are generated with the LO \madgraph\ generator,
we add an asymmetry of  9.5\% arising from the pure SM contributions
that is not accounted for by \madgraph.
The approximation of just adding the \madgraph\ LO asymmetry to the SM asymmetry is estimated to be valid at the $\approx 3\%$ level.
 
\invisible{
The results are presented in Fig.~\ref{fig:2d_result_afb_polar},
where we overlay the expected values for the different axigluon models of  Ref.~\cite{Carmona:2014gra}.
As the models are generated with the LO \madgraph\ generator,
we add an asymmetry of  9.5\%~\cite{Czakon:2014xsa} arising from the pure SM contributions
that is not accounted for by \madgraph.
The approximation of just adding the \madgraph\ LO asymmetry to the SM asymmetry is estimated to be valid at the $\approx 3\%$ level.
The NLO SM prediction for \Attfb\ is $\Att =  (9.5\pm 0.7)\% $~\cite{Czakon:2014xsa}, while  the SM polarization is expected to be small,  $\Ptt=(-0.19\pm 0.05)\% $~\cite{Bernreuther:2006vg}. Our measurement is consistent with the SM prediction within the 68\% confidence level region.
}

We interpret the measurements as a test of the SM, separately assuming 
the SM forward-backward asymmetry of  $\Att=(9.5\pm0.7) \%$
and the SM polarization of $\Ptt= (-0.19 \pm 0.05) \%$.
As we assume the SM, we do not consider the
uncertainty from the dependence on the physics model.
The  constraint on \Att\ is applied to the two-dimensional result of Eq.~(\ref{eq:final_result}) to obtain the polarization
\begin{eqnarray}
\Ptt &=& (11.3 \pm 9.1 \text{ (stat)}  \pm 1.9 \text{ (syst)})\%.
\label{eq:Ptt_only_result}  \end{eqnarray}
This result is consistent with the SM expectation at the 1.2 standard deviation level.
Applying  the constraint on \Ptt\ we  obtain an asymmetry of
\begin{eqnarray}
\Att &=& (17.5 \pm 5.6 \text{ (stat)}  \pm 3.1 \text{ (syst)})\%,
\label{eq:Att_only_result}  \end{eqnarray}
which is consistent with  the SM expectation at the 1.3 standard deviation level.

In a previous publication, the \dzero\ Collaboration has measured the forward-backward
asymmetry in the
lepton+jets channel~\cite{Abazov:2014cca}: \begin{equation}
\Att = (10.6\pm 2.8  \text{ (stat)}   \pm 1.3 \text{ (syst)})\%= (10.6\pm3.0)\% .\end{equation}
This lepton+jets measurement  was performed in the context of a test of the SM, as no study of the dependence with respect to the possible polarization was performed. Therefore, it should be compared with the result of Eq.~(\ref{eq:Att_only_result}).
We  classify the systematic uncertainties of both measurements by their sources and consider them as being
either completely correlated, \eg, the $b$-tagging uncertainty, or completely uncorrelated, \eg, the background modeling.
Even if some sources of uncertainties are correlated between both channels, the dominant sources are not, so that the final overall uncertainties are only $7\%$ correlated.
The two measurements are consistent with a  probability of 30\% given by a $\chi^2$ test.
We combine the lepton+jets and dilepton measurements, using the best linear unbiased estimate (BLUE)~\cite{Lyons:1988rp,Valassi:2003mu}.
The combination is a weighted average of the
input measurements, with the dilepton measurement given
a weight of 0.17 and the lepton+jets measurement a weight of 0.83.
The combined result is
\begin{eqnarray}
\Att &=& 11.8 \pm 2.5  \text{ (stat)}   \pm 1.3 \text{ (syst)})\% .
\end{eqnarray}

\section{Summary}

We have presented a simultaneous measurement of
the forward-backward asymmetry of \ttbar\ production and the top quark spin polarization in the beam basis in
dilepton final states, using \lumi\ of
proton-antiproton collisions at $\sqrt{s}=1.96$~TeV with the D0 detector.
The results are:
\begin{eqnarray}
\Att &=& (15.0 \pm 8.0)\%, \label{eq:final_result2}\\
\Ptt &=& (7.2 \pm 11.3)\%, \nonumber
\end{eqnarray}
with a correlation of $-56\%$ between the  measurements. They are consistent with the SM expectations within the 68\% confidence level region.

Interpreted as a test of the SM and assuming the SM forward-backward asymmetry,  these results yield  a measurement of the top polarization of
\begin{eqnarray}
\Ptt &=& (11.3 \pm 9.3)\%. 
\end{eqnarray}

Assuming the SM   polarization, we obtain a forward-backward asymmetry of
\begin{eqnarray}
\Att &=& (17.5 \pm  6.3)\%. 
\end{eqnarray}
This asymmetry is combined with the measurement of the asymmetry in lepton+jets final states yielding a combined asymmetry of
\begin{eqnarray}
\Att &=& (11.8 \pm 2.8)\%.
\end{eqnarray}

All of these results are consistent with the SM expectations within  uncertainties.

\section{\label{sec:ackn}ACKNOWLEDGMENTS}
\makeatletter{}
We thank the staffs at Fermilab and collaborating institutions,
and acknowledge support from the
Department of Energy and National Science Foundation (United States of America);
Alternative Energies and Atomic Energy Commission and
National Center for Scientific Research/National Institute of Nuclear and Particle Physics  (France);
Ministry of Education and Science of the Russian Federation, 
National Research Center ``Kurchatov Institute" of the Russian Federation, and 
Russian Foundation for Basic Research  (Russia);
National Council for the Development of Science and Technology and
Carlos Chagas Filho Foundation for the Support of Research in the State of Rio de Janeiro (Brazil);
Department of Atomic Energy and Department of Science and Technology (India);
Administrative Department of Science, Technology and Innovation (Colombia);
National Council of Science and Technology (Mexico);
National Research Foundation of Korea (Korea);
Foundation for Fundamental Research on Matter (The Netherlands);
Science and Technology Facilities Council and The Royal Society (United Kingdom);
Ministry of Education, Youth and Sports (Czech Republic);
Bundesministerium f\"{u}r Bildung und Forschung (Federal Ministry of Education and Research) and 
Deutsche Forschungsgemeinschaft (German Research Foundation) (Germany);
Science Foundation Ireland (Ireland);
Swedish Research Council (Sweden);
China Academy of Sciences and National Natural Science Foundation of China (China);
and
Ministry of Education and Science of Ukraine (Ukraine).

\bibliographystyle{apsrev_custom2}
\bibliography{references}

\end{document}